\documentclass[a4paper,fleqn]{cas-dc}

\usepackage[numbers]{natbib}

\usepackage{amsmath,amssymb,amsfonts}
\usepackage{algorithmic}
\usepackage{graphicx}
\usepackage{textcomp}

\usepackage{tabularx}

\usepackage{multirow}

\def\tsc#1{\csdef{#1}{\textsc{\lowercase{#1}}\xspace}}
\tsc{WGM}
\tsc{QE}

\begin{document}
\let\WriteBookmarks\relax
\def\floatpagepagefraction{1}
\def\textpagefraction{.001}

\shorttitle{A Low-Complexity View Synthesis Distortion Estimation Method}    

\shortauthors{C. Bi \& J. Liang}  

\title [mode = title]{A Low-Complexity View Synthesis Distortion Estimation Method for 3D Video with Large Baseline Considerations}  

\tnotemark[1] 

\tnotetext[1]{This paper is supported by Natural Sciences and Engineering Research Council of Canada (NSERC) grant RGPIN-2020-04525.} 

%

\author{Chongyuan Bi}[role=Student]



\ead{cba30@sfu.ca}

\author{Jie Liang}[role=Professor]

\cormark[1]

\ead{jiel@sfu.ca}

\affiliation{organization={School of Engineering Science, Simon Fraser University},
            addressline={8888 University Drive}, 
            city={Burnaby},
            postcode={V5A 1S6}, 
            state={BC},
            country={Canada}}




\begin{abstract}
Depth-image-based rendering is a key view synthesis algorithm in 3D video systems, which enables the synthesis of virtual views from texture images and depth maps. An efficient view synthesis distortion estimation model is critical for optimizing resource allocation in real-time applications such as interactive free-viewpoint video and 3D video streaming services. However, existing estimation methods are often computationally intensive, require parameter training, or performance poorly in challenging large baseline configurations. This paper presents a novel, low-complexity, and training-free method to accurately estimate the distortion of synthesized views without performing the actual rendering process. Key contributions include: (1) A joint texture-depth classification method that accurately separates texture image into locally stationary and non-stationary regions, which mitigates misclassifications by using texture-only methods. (2) A novel baseline distance indicator is designed for the compensation scheme for distortions caused by large baseline configurations. (3) A region-based blending estimation strategy that geometrically identifies overlapping, single-view, and mutual disocclusion regions, predicting distortion in synthesized views from two reference views with differing synthesis conditions. Experiments on standard MPEG 3D video sequences validate the proposed method's high accuracy and efficiency, especially in large baseline configurations. This method enables more flexible camera arrangements in 3D content acquisition by accurately predicting synthesis quality under challenging geometric configurations.
\end{abstract}


\begin{highlights}
\item This paper presents a novel, low-complexity, and training-free method to accurately estimate the distortion of synthesized views without performing the actual rendering process.
\item A joint texture-depth classification method that accurately separates texture image into locally stationary and non-stationary regions, which mitigates misclassifications by using texture-only methods.
\item A novel baseline distance indicator is designed for the compensation scheme for distortions caused by large baseline configurations.
\item An accurate spatial-domain method is designed to estimate VSD in NS regions, improving prediction accuracy for large baselines while maintaining low complexity.
\item A region-based blending estimation strategy that geometrically identifies overlapping, single-view, and mutual disocclusion regions, predicting distortion in synthesized views from two reference views with differing synthesis conditions.
\end{highlights}


\begin{keywords}
3D video \sep baseline distance indicator \sep depth-image-based rendering \sep large baseline \sep low-complexity \sep view synthesis distortion estimation.
\end{keywords}

\maketitle

\section{Introduction}
\label{sec:introduction}

\subsection{Motivation}

3D video (3DV) technology enables immersive viewing experiences such as Free-Viewpoint Video (FVV) and 3D television (3D-TV). Multiview Video plus Depth (MVD) format is a popular format in these systems, where a texture video is associated with a per-pixel depth map \cite{2004SPIE_Fehn}. This combination of texture and depth data is referred to as a reference view. New virtual views can then be generated at user-chosen viewpoints through Depth-Image-Based Rendering (DIBR) algorithm \cite{2004SPIE_Fehn}, which uses one or multiple reference views as input.

While the MVD representation can significantly reduce the data size of 3DV, it still produces a substantial amount of data. Consequently, bandwidth and storage constraints typically limit transmission to only a few compressed reference views. This requires the selection of optimal reference views and an optimal bit allocation strategy between the texture and depth videos to maximize the final synthesized view quality. The core challenge lies in synthesized view distortion estimation. Because users do not view depth maps directly, traditional video coding metrics that assess texture and depth distortions independently are suboptimal for this task. Therefore, a specialized View Synthesis Distortion Estimation (VSDE) method is required at the encoder to accurately predict the final distortion at the decoder.

For real-time applications such as interactive FVV and 3DV streaming, a low-complexity VSDE method is particularly critical. Many existing VSDE approaches are computationally intensive, as they either require the entire DIBR rendering process or involve extensive, content-specific parameter training. Furthermore, the failure of many current models to account for distortion from large baselines makes them unsuitable for many practical applications.

\subsection{Challenges with Existing Methods}

As we discussed earlier, FVV and 3D streaming services require low-complexity and accurate VSDE method to ensure good user quality of service (QoS). However, existing VSDE methods face several challenges that limit their use in practice.

The method in \cite{2011TIP_Gene} develops a cubic distortion model and a 3D trellis algorithm for bit allocation in multiview image coding. \cite{2014TCSVT_Yuan} derives a polynomial distortion model to quantify the impact of depth map coding errors on synthesized virtual views in 3DV systems and uses rate-distortion optimization to improve depth map coding efficiency. \cite{2016TIP_Fang} proposes an analytical model to estimate depth-error-induced Virtual View Synthesis Distortion (VVSD) in 3DV. It classifies VVSD into nine scenarios and models the distortion in different scenarios with quadratic or biquadratic functions, while modeling their probabilities with linear functions of the virtual view position. However, all these methods require significant parameter training or fitting. \cite{2015TCSVT_Dong} introduces a probabilistic graphical model and a Recursive Optimal Distribution Estimation (RODE) method to accurately estimate decoder-side distortion of synthesized virtual views. \cite{2022TMM_Jin} proposed an auto-weighted layer representation model for pixel-level VSDE in 3DV coding. While these methods are accurate, they are computationally expensive due to pixel-level processing. \cite{2019TMM_Meng} develops an efficient and parameter-free VSDE scheme for depth coding optimization, integrating it into Rate-Distortion Optimization (RDO). However, it only considers one reference view, which limits its applicability in complex multiview scenarios. Deep learning approaches, such as the quality assessment model by \cite{2024ASC_Shi}, require large pre-training datasets. \cite{2016TIP_Yu} proposes an encoder-driven inpainting strategy for multiview video compression that uses DIBR to synthesize views and employs template-matching to fill disocclusion holes, but requires complex auxiliary information transmission and iterative optimization processes that increase computational overhead. The method of \cite{2014TIP_Fang} estimates synthesized view quality by analyzing texture and depth induced distortions in different domains. It still face limitations in challenging situations, such as large baseline configurations. There are also many studies focusing on subjective quality assessment, such as those by \cite{2019TIP_Dragana} and \cite{2020TIP_Wang}, but they are not suitable for RDO in practical applications.

In practical 3DV applications, cameras are usually not densely placed due to resource constraints. As a result, large baselines between reference views are common. This introduces stronger geometric disparities, larger disoccluded regions, and amplifies the impact of depth errors on the synthesized view distortion. Existing VSDE methods often fail to account for these effects under large baseline configurations. \cite{2012TMM_Xiaoyu} addresses interactive multiview streaming with free viewpoint synthesis using joint optimization of frame structure and quantization parameters, but assumes simplified network conditions and does not specifically address the distortion estimation challenges that arise from large baseline configurations.

\subsection{Overview of the Proposed Method and Contributions}

This paper introduces a novel, low-complexity, and training-free VSDE method to estimate the Mean Squared Error (MSE) of synthesized views without requiring the actual DIBR process. A key focus is to accurately account for distortions arising from large baseline configurations between reference and virtual views, which is a common challenge in practical 3DV applications. The method assumes a 1D parallel rectified linear camera setup.

The main contributions are as follows.
\begin{itemize}
    \item Proposing a joint texture-depth classification method that improves the accuracy of separating Locally Stationary (LS) and Non-Stationary (NS) regions.
    \item Developing a refined VSDE model for LS regions that integrates a novel Baseline Distance Indicator (BDI) and a compensation mechanism to more accurately predict View Synthesis Distortion (VSD) arising from large baseline configurations.
    \item Designing an accurate spatial-domain method to estimate VSD in NS regions, improving prediction accuracy for large baselines while maintaining low complexity.
    \item Developing an effective geometric region-based blending estimation strategy for estimating the blended VSD when using two reference views, accounting for overlapping regions, single-view regions, and mutual disocclusions that become significant in large baseline configurations.
\end{itemize}

By successfully addressing VSDE for large baseline configurations, this method facilitates more flexible and sparse camera arrangements in 3DV acquisition. And it can potentially reduce the related costs and complexity without compromising the quality of synthesized views.

\subsection{Structure of the Paper}

The remainder of this paper is organized as follows: Section~\ref{sec:system_model} describes the system model, VSD decomposition, and analysis of challenges due to large baseline configurations. Section~\ref{sec:VSDE} presents the proposed VSDE method in detail, beginning with the model for VSD caused by texture coding. It then describes the comprehensive framework for VSD caused by depth coding, which includes the joint texture-depth classification method, specialized estimation models for LS and NS regions, and an effective strategy for estimating blended distortion, all with a key focus on addressing large baseline configurations. Section~\ref{sec:experiments} conducts simulations for the validation of the proposed method. Finally, Section~\ref{sec:conclusion} concludes the paper and discusses the future research plan.

\section{System Model and Analysis}
\label{sec:system_model}

This section establishes the foundation for our proposed VSDE method. We first introduce the system model. Next, we analyze the VSD using a mathematical framework that decomposes the effects of coding errors. Finally, we analyze the specific challenges that arise when synthesizing virtual views across large baseline configurations.

\begin{table}[h]
    \centering
    \caption{Notations}
    \label{tab:notations}
    \begin{tabularx}{\columnwidth}{lX}
        \hline
        \textbf{Notation} & \textbf{Definition} \\
        \hline
        $T_L$, $T_R$ & Original reference texture images captured from the left and right cameras respectively \\
        $D_L$, $D_R$ & Original depth maps associated with the left and right views respectively \\
        $\hat{T_L}$, $\hat{T_R}$ & Compressed/reconstructed texture images for the left and right cameras respectively \\
        $\hat{D_L}$, $\hat{D_R}$ & Compressed/reconstructed depth maps associated with the left and right views respectively \\
        $U_L$, $U_R$ & Left and right intermediate virtual views rendered from warping $T_L$ with depth map $D_L$, and $T_R$ with depth map $D_R$ respectively \\
        $U$ & Synthesized virtual view at the virtual position using $T_L$, $T_R$, $D_L$, $D_R$ \\
        $Y_L$, $Y_R$ & Left and right intermediate virtual views rendered from warping $\hat{T_L}$ with depth map $D_L$, and $\hat{T_R}$ with depth map $D_R$ respectively \\
        $Y$ & Synthesized virtual view at the virtual position using $\hat{T_L}$, $\hat{T_R}$, $D_L$, $D_R$ \\
        $W_L$, $W_R$ & Left and right intermediate virtual views rendered from warping $\hat{T_L}$ with depth map $\hat{D_L}$, and $\hat{T_R}$ with depth map $\hat{D}_R$ respectively \\
        $W$ & Synthesized virtual view at the virtual position using $\hat{T_L}$, $\hat{T_R}$, $\hat{D_L}$, $\hat{D_R}$ \\
        $N$ & VSD caused by texture images coding, $N = U - Y$ \\
        $Z$ & VSD caused by depth maps coding, $Z = Y - W$ \\
        $V$ & Overall VSD caused by texture images and depth maps coding, $V=U-W$ \\
        \hline
    \end{tabularx}
\end{table}

\subsection{System Model}
\label{subsec:system_model}

DIBR is the standard algorithm for synthesizing virtual views from reference texture images and depth maps \cite{2009SPIE_Dong, 2011TCSVT_Horng}. In this work, we focus on the 1D parallel rectified camera arrangement used in MPEG 3DV experiments, where cameras are linearly arranged with parallel optical axes. This configuration simplifies the warping process to a simple horizontal disparity shift. The system model is designed for estimating the total VSD in MSE metric that results from using reconstructed texture images and depth maps from two reference views, without performing the computational intensive DIBR process.

The system takes the following data as inputs:
\begin{itemize}
    \item Original reference views: Two pairs of uncompressed texture images $T_L,T_R$ and their corresponding depth maps $D_L,D_R$.
    \item Reconstructed reference views: The corresponding two pairs of reconstructed texture images $\hat{T_L},\hat{T_R}$ and depth maps $\hat{D_L},\hat{D_R}$ after compression.
    \item Camera parameters: The intrinsic and extrinsic parameters of the camera setup, which are necessary for the geometric calculations in the view synthesis process.
\end{itemize}

Our goal is to estimate the VSD between virtual views synthesized from the original versus reconstructed reference views. Following \cite{2014TCSVT_Yuan} and \cite{2014TIP_Fang}, we decompose the total VSD into separate contributions from texture and depth coding errors. Table~\ref{tab:notations} summarizes the notations used throughput this paper, where $V=U-W$ represents the total VSD between the ground truth virtual view $U$ (synthesized from original data) and the distorted virtual view $W$ (synthesized from compressed data).

\subsection{Analysis of Coding Errors Impact on the Synthesized Virtual View}
\label{subsec:analysis_coding_error_impact}

In the view synthesis pipeline, firstly, two original texture images $T_L$ and $T_R$ along with their original depth maps $D_L$ and $D_R$ are used to generate the intermediate virtual view $U_L$ and $U_R$. Specifically, in the warping step, pixels from $T_L(x',y)$ are copied to $U_L(x,y)$ to form the left intermediate virtual view $U_L$, where $(x',y)$ and $(x,y)$ are the pixel coordinates. There is only horizontal disparity due to the assumption of 1D parallel rectified linear camera setup. And the horizontal disparity is given by \cite{2009VCIP_Lai}

\begin{equation}
    \begin{split}
    x-x' & = k_L \cdot D_L(x',y) + c \\
    \text{where } & k_L = \frac{f \cdot b_L}{255}(\frac{1}{Z_{near}}-\frac{1}{Z_{far}}) \\
    & c = \frac{f \cdot b_L}{Z_{far}}
    \end{split}
    \label{eq:horizontal_disparity}
\end{equation}

In this equation, $f$ is the focal length of the camera, $b_L$ is the baseline distance between left reference view and the virtual view, and $Z_{near}$ and $Z_{far}$ are the nearest and farthest depth values of the scene. 

Similarly, pixels from $T_R(x'',y)$ are copied to $U_R(x,y)$ to form the right intermediate virtual view $U_R$. Then, $U_L$ and $U_R$ are blended to generate the final virtual view $U$ by commonly used linear combination as

\begin{equation}
    U(x,y)=\alpha U_L(x,y)+(1-\alpha)U_R(x,y)
    \label{eq:linear_combination}
\end{equation}

\begin{equation}
    \alpha = \frac{b_R}{b}
    \label{eq:blending_weight}
\end{equation}

$\alpha$ is the linear combination weight, $b_L$ is the baseline distance between the left reference view and the virtual view, $b_R$ is the baseline distance between left reference and the virtual views, and $b=b_L+b_R$ is the baseline distance between the left and right reference views. After that, any remaining holes are filled using information from neighboring pixels or more advanced techniques like inpainting.

$U$ acts as the ground truth virtual view for VSD computation. Similarly, two reconstructed reference texture images $\hat{T_L}, \hat{T_R}$ along with their original depth maps $D_L, D_R$ are used to generate the synthesized virtual view $Y$. Since $U$ and $Y$ are only different due to the reconstructed texture images $\hat{T_L}, \hat{T_R}$, $N=U-Y$ is the distortion due to the compression of texture images. Finally, two reconstructed reference texture images $\hat{T_L}, \hat{T_R}$ along with their reconstructed depth maps $\hat{D_L}, \hat{D_R}$ are used to generate the synthesized virtual view $W$. Thus, $Z=Y-W$ represents the additional distortion from depth compression, and $V=N+Z$ is the total VSD.

Following \cite{2014TCSVT_Yuan}, the total expected squared distortion can be expressed as:

\begin{equation}
    \begin{split}
    E[V^2]& = E_{tex}[N^2]+E_{dep}[Z^2]+E[NZ] \\
    & \approx E_{tex}[N^2] + E_{dep}[Z^2]
    \label{eq:VSD_separation}
    \end{split}
\end{equation}

The approximation holds because quantization errors from texture and depth compression are assumed to be independent zero-mean white noises, making their cross-correlation $E[NZ]$ approximately zero \cite{2014TCSVT_Yuan}. This decomposition enables separate estimation of texture compression distortion ($E_{tex}[N^2]$) and depth compression distortion ($E_{dep}[Z^2]$), significantly simplifying the VSDE analysis. Further details can be found in \cite{2014TCSVT_Yuan} and \cite{2014TIP_Fang}.

In practice, analyzing VSD due to depth map coding is significantly more complex than VSD due to texture coding, as depth errors cause geometric warping artifacts rather than simple pixel value differences. Addressing this complexity is the central focus of the following sections. 

The depth coding error for the left reference view is defined as:
\begin{equation}
    \Delta D_L(x,y) = \hat{D_L}(x,y) - D_L(x,y)
    \label{eq:depth_error}
\end{equation}

Following \cite{2009VCIP_Lai}, this depth error causes a proportional horizontal disparity error:
\begin{equation}
    \Delta x_L(x,y) = k_L \cdot \Delta D_L(x,y)
    \label{eq:disparity_error}
\end{equation}

where $k_L$ is defined in equation \eqref{eq:horizontal_disparity}. Consequently, a pixel that should be warped from position $x'$ in the reference view is instead warped from position $x'+\Delta x_L$ due to the depth error. The resulting VSD at each pixel is:
\begin{equation}
    Z_L(x,y) = \hat{T_L}(x'+\Delta x_L,y)-\hat{T_L}(x',y)
    \label{eq:vsd_depth_error}
\end{equation}

\subsection{Large Baseline Considerations}
\label{subsec:large_baseline_considerations}

While existing VSDE models provide valuable frameworks \cite{2014TCSVT_Yuan,2014TIP_Fang,2019TMM_Meng}, from detailed analytical approaches to low-complexity parameter-free methods, they all have a common limitation: the assumption of small baseline distances between reference and virtual views. This assumption fails in practical applications where bandwidth constraints require sparse camera arrangements or limit the number of transmitted reference views. Therefore, prior models struggle to accurately estimate VSD in large baseline configurations.

The primary challenges associated with large baselines include:

\begin{itemize}
    \item Error amplification: Larger baselines cause greater pixel position shifts during warping. From \eqref{eq:horizontal_disparity} and \eqref{eq:disparity_error}, the disparity error $\Delta x_L(x,y)$ is proportional to the baseline distance $b_L$. Consequently, small depth inaccuracies could produce disproportionately large geometric distortions in the synthesized view, significantly increasing VSD sensitivity to depth coding errors.
    
    \item Significant disocclusions: As the virtual viewpoint moves further from reference views, disocclusion regions expand dramatically in both size and frequency. These regions create substantial challenges for accurate distortion estimation since no information from reference views can be used to fill them.
    
    \item Blending and hole-filling complexity: Multi-view blending becomes increasingly difficult with large baselines due to inconsistent warping artifacts between views. Similarly, hole-filling algorithms struggle with the expanded disocclusion regions characteristic of large baseline synthesis.
\end{itemize}

To address these limitations, our proposed method evaluates the severity of large baseline effects for given scene and camera configurations, then applies a novel compensation model to accurately estimate VSD in these challenging configurations. This model maintains the computational efficiency and parameter-free advantages of existing methods. Furthermore, our method addresses their primary weakness: the inability to handle large baseline configurations effectively.

\section{Overall Framework of the Proposed VSDE Method}
\label{sec:VSDE}

As discussed in Section~\ref{subsec:analysis_coding_error_impact}, the overall VSDE $E[V^2]$ can be decomposed into texture coding caused VSDE $E_{tex}[N^2]$ and depth coding caused VSDE $E_{dep}[Z^2]$. This section presents the complete framework for estimating these components.

\subsection{Overall VSDE Pipeline}
\label{subsec:overall_VSDE_pipeline}

As described in Section~\ref{subsec:system_model}, our proposed VSDE method takes the original and reconstructed reference views and camera parameters as inputs. As shown in Fig.~\ref{fig:flowchart}, the main steps of the overall VSDE pipeline are as follows:

\begin{figure}
    \centering
    \includegraphics[width=1\linewidth]{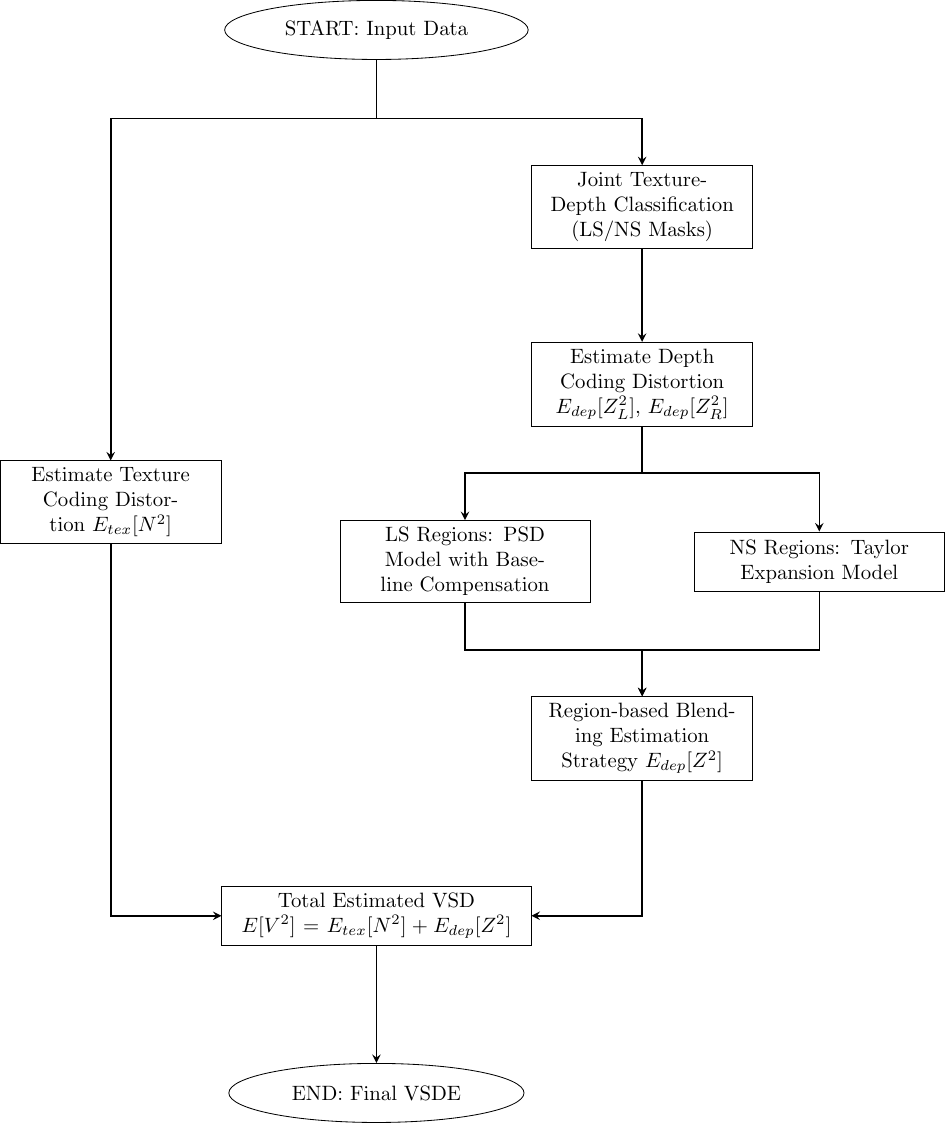}
    \caption{VSDE pipeline}
    \label{fig:flowchart}
\end{figure}

\begin{enumerate}
    \item Estimate texture coding caused VSD $E_{tex}[N^2]$: The VSD caused by texture coding is calculated under the assumption of original depth maps.
    \item Perform joint texture-depth classification: The reference texture frames are segmented into LS and NS regions using a joint texture-depth classification method to generate LS/NS masks.
    \item Estimate depth coding caused VSD $E_{dep}[Z_L^2]$ and $E_{dep}[Z_R^2]$: For left and right reference views, the VSDs caused by depth coding are estimated separately for LS and NS regions.
    \begin{itemize}
        \item For LS regions, VSDs $E_{LS}[Z_L^2]$ and $E_{LS}[Z_R^2]$ are estimated in frequency-domain using a Power Spectral Density (PSD) model, which is then adjusted by a novel large baseline compensation mechanism.
        \item For NS regions, VSDs $E_{NS}[Z_L^2]$ and $E_{NS}[Z_R^2]$ are estimated using a Taylor expansion spatial-domain model.
    \end{itemize}
    \item Estimate blended depth coding caused VSD $E_{dep}[Z^2]$: The estimated depth coding VSDs from the left and right reference views are combined using a geometric region-based blending estimation strategy to produce the final depth coding caused VSDE. This strategy accounts for overlapping regions, single-view regions, and mutual disocclusion regions.
    \item Calculate total estimated VSD $E[V^2]$: The final estimated VSD is the sum of the estimated VSDs caused by texture coding and depth coding.
\end{enumerate}

\subsection{Estimation of VSD Caused by Texture Coding}
\label{subsec:texture_VSDE}

Following the approaches of \cite{2014TCSVT_Yuan} and \cite{2014TIP_Fang}, the VSD caused by texture coding can be estimated under the assumption of original depth maps as:

\begin{equation}
    \begin{split}
    E_{tex}[N^2] = {} & \alpha^2E[(T_L-\hat{T_L})^2]+(1-\alpha)^2E[(T_R-\hat{T_R})^2] \\
    & +2\alpha(1-\alpha) \rho_N \sigma_{T_L-\hat{T_L}} \sigma_{T_R-\hat{T_R}}
    \end{split}
    \label{eq:texture_VSD}
\end{equation}

where $E[(T_i-\hat{T_i})^2]$ $(i=L,R)$ represents the MSE of the texture coding error, $\rho_N$ is the correlation coefficient between texture coding errors from left and right reference views, $\sigma_{T_i-\hat{T_i}}$ is the corresponding standard deviation, and $\alpha$ is the blending weight from \eqref{eq:blending_weight}. 

We assume $\rho_N \approx 0$ for independently encoded views. This assumption is reasonable when compression artifacts are mainly random noise rather than strong structural correlations. Thus, \eqref{eq:texture_VSD} simplifies to:

\begin{equation}
    E_{tex}[N^2] \approx \alpha^2E[(T_L-\hat{T_L})^2]+(1-\alpha)^2E[(T_R-\hat{T_R})^2]
    \label{eq:texture_VSD_simplified}
\end{equation}

Note that $\rho_N$ cannot be ignored if the texture images are encoded at very low quality. In this case, $(T_L-\hat{T_L})$ and $(T_R-\hat{T_R})$ would be more correlated, and a model needs to be trained to estimate $\rho_N$. Details can be found in \cite{2014TIP_Fang}.

\subsection{Joint Texture-Depth Classification for LS/NS Separation}
\label{subsec:joint_texture_depth_classification}

Accurate classification of image regions into LS and NS categories is crucial for our method, as we employ different VSDE models for these regions. This classification is performed on the reference views before warping and does not include disoccluded regions, which appear only during the synthesis process and are handled separately through hole-filling (Section~\ref{subsec:estimation_blended_VSD}). LS regions are areas with relatively homogeneous texture or smoothly varying content that can be modeled as locally ergodic Wide-Sense Stationary (WSS) processes, where warping generates mostly uniform distortion patterns. In contrast, NS regions occur around object edges, depth discontinuities, or areas with rapidly changing statistical properties that violate the WSS assumption, resulting in substantial and spatially varying synthesis distortions.

In \cite{2014TIP_Fang}, the LS/NS segmentation was based only on texture image gradients. However, relying only on the texture image can misclassify regions. For example, a highly textured region co-located with a smooth depth region could be misclassified as NS, while a smooth texture region co-located with a sharp depth edge could be misclassified as LS. The proposed joint texture-depth classification method addresses this issue through the following steps:

\begin{enumerate}
    \item Gradient computation: We compute gradient maps for both the original texture image and depth map of each reference view using the Sobel operator. Let $G_{Ti}(x,y)$ and $G_{Di}(x,y)$ denote the texture and depth gradient magnitudes at pixel $(x,y)$ for reference view $i \in \{L,R\}$.
    
    \item Joint Edge Map (JEM) formulation: Each gradient map is normalized to $[0,1]$ to ensure comparable scales:
    
    \begin{equation}
        \begin{split}
            G_{Ti,norm}(x,y) &= \frac{G_{Ti}(x,y)-\min_{(x,y)}(G_{Ti})}{\max_{(x,y)}(G_{Ti})-\min_{(x,y)}(G_{Ti})} \\
            G_{Di,norm}(x,y) &= \frac{G_{Di}(x,y)-\min_{(x,y)}(G_{Di})}{\max_{(x,y)}(G_{Di})-\min_{(x,y)}(G_{Di})}
        \end{split}
    \end{equation}
    
    where the min and max operations are computed over all pixels in the respective frames.
    
    The JEM combines these normalized gradients as follows (pixel coordinates omitted for clarity):
    
    \begin{equation}
        G_{JEM} = w_{D} \cdot G_{D,norm} + w_{T} \cdot G_{T,norm} \cdot (1 - G_{D,norm})
        \label{eq:JEM}
    \end{equation}
    
    where $w_{D}$ and $w_{T}$ are the weights for depth and texture gradients, respectively. For this low-complexity model, we set $w_{D}=1$ and $w_{T}=0.5$ heuristically. This formulation prioritizes depth gradients, as sharp depth discontinuities are the primary cause of significant synthesis artifacts like disocclusions. The term $(1 - G_{D,norm})$ attenuates texture gradient contributions in areas where strong depth gradients already indicate NS regions, thus preventing misclassification.
    
    \item Thresholding for LS/NS separation: We apply Otsu's method to automatically threshold $G_{JEM}(x,y)$ and segment LS/NS regions. This method is chosen for its parameter-free nature and effectiveness in separating bimodal histograms. Pixels with JEM values above the Otsu-determined threshold are labeled as NS, while the remaining pixels are labeled as LS. This process is performed independently for both left and right reference views, generating separate LS/NS masks that are subsequently used in the region-specific VSD estimation steps.
\end{enumerate}

\subsection{Estimation of VSD Caused by Depth Coding in LS Regions}
\label{sec:VSDE_LS}

In this section, we study the VSD caused by depth coding in LS regions. These regions exhibit spatially uniform statistical properties and can be modeled as realizations of an ergodic WSS random field.

\subsubsection{Basic PSD Model}
\label{subsubsec:basic_PSD_model}

Following the methods of \cite{2014TIP_Fang} and \cite{1987JSAC_Girod}, we estimate the VSD in LS regions of the left reference view using a PSD based model in the frequency-domain:

\begin{equation}
    \Phi_{Z_{L,LS}}(\omega_1,\omega_2) \approx 2(1-\text{Re}\{ P_L(\omega_1) \}) \Phi_{\hat{T}_{L,LS}}(\omega_1,\omega_2)
    \label{eq:PSD_basic}
\end{equation}

where $\Phi_{Z_{L,LS}}(\omega_1,\omega_2)$ is the PSD of the depth coding distortion in LS regions of the left view, $P_L(\omega_1)$ is the Fourier Transform (FT) of the probability distribution of horizontal disparity errors $p(\Delta x_L)$ for the left view, and $\Phi_{\hat{T}_{L,LS}}(\omega_1,\omega_2)$ is the PSD of the reconstructed texture in LS regions of the left reference view.

Under small baseline configurations, this PSD-based model provides accurate VSDE. The expected distortion in LS regions is obtained by integrating the PSD over all spatial frequencies:

\begin{equation}
    E_{LS,PSD}[Z_L^2] = \frac{1}{(2\pi)^2} \int_{-\pi}^{\pi} \int_{-\pi}^{\pi} \Phi_{Z_{L,LS}}(\omega_1,\omega_2) d\omega_1 d\omega_2
    \label{eq:PSD_integration}
\end{equation}

However, for large baseline configurations, this basic model tends to underestimate the actual distortion due to factors not captured in the frequency-domain analysis, such as increased disocclusions and amplified warping artifacts. These limitations motivate the compensation mechanism presented in the following sections.

\subsubsection{Baseline Distance Indicator Design}
\label{subsubsec:BDI}

To address the limitations identified above, we introduce a baseline compensation framework based on the BDI. The BDI is a scalar metric that quantitatively measures the severity of the baseline configuration by combining four key factors that increase VSD as baseline increases. Each factor is normalized to [0,1] and incorporated into a weighted sum:

\begin{enumerate}
    \item Physical baseline distance $D_{phys,norm}$: The normalized physical baseline distance between the reference and virtual views. For the left reference, $D_{phys} = b_L$, and for the right reference, $D_{phys} = b_R = b - b_L$. This factor directly influences parallax: larger baseline distances yield greater disparity shifts. The normalization uses the nearest scene depth value $Z_{near}$, since the nearest objects produce the largest disparities:

    \begin{equation}
        D_{phys,norm} = \min\left(\frac{D_{phys}}{Z_{near}}, 1\right)
    \end{equation}

    where both the baseline distance $D_{phys}$ and $Z_{near}$ are expressed in the same physical units (e.g., meters).

    \item Estimated disocclusion area $O_{disocc,norm}$: The normalized estimated disocclusion area, crucial for VSDE in large baseline configurations where significant disocclusion occurs. The estimation method consists of three steps:

    \begin{itemize}
        \item Step 1: Detect significant depth discontinuities: Apply the Sobel operator to the reference depth map to detect depth discontinuities (reusing results from Section~\ref{subsec:joint_texture_depth_classification}). This produces a depth gradient map $G_d$, from which a binary edge mask $M_{edge}$ is obtained by thresholding.
        
        \item Step 2: Estimate disocclusion width at each edge pixel: For each pixel in $M_{edge}$, estimate the width of the disocclusion that will occur during view synthesis. This width equals the disparity difference between foreground and adjacent background pixels:

        \begin{equation}
            W_{disocc}(x,y) = |d_{fg} - d_{bg}|
        \end{equation}

        Substituting the disparity formula from \cite{2009SPIE_Dong}:

        \begin{equation}
            W_{disocc}(x,y) = f \cdot b_i \left|\frac{1}{z_{fg}} - \frac{1}{z_{bg}}\right|
        \end{equation}

        where $f$ is the focal length, $b_i$ $(i \in \{L,R\})$ is the baseline distance from the reference to virtual view, $z_{fg}$ and $z_{bg}$ are the physical depth values of the foreground and background pixels. These are computed from the depth map values $D(x,y) \in [0,255]$ as:

        \begin{equation}
            z = \frac{1}{\frac{D(x,y)}{255}\left(\frac{1}{Z_{near}} - \frac{1}{Z_{far}}\right) + \frac{1}{Z_{far}}}
        \end{equation}
        
        \item Step 3: Calculate and normalize total disocclusion area: Sum all individual disocclusion widths and normalize by frame size:

        \begin{equation}
            O_{disocc} = \sum_{(x,y) \in M_{edge}} W_{disocc}(x,y)
        \end{equation}

        \begin{equation}
            O_{disocc,norm} = \frac{O_{disocc}}{W \times H}
            \label{eq:disocclusion_factor}
        \end{equation}

        where $W$ and $H$ are the frame width and height.
    \end{itemize}
    
    \item Maximum disparity $D_{max\_disp,norm}$: The maximum disparity, indicating the largest pixel displacement during synthesis. Computed from \eqref{eq:horizontal_disparity} and normalized by image width $W$.
    
    \item Texture complexity $T_{comp,norm}$: The average gradient magnitude across the texture frame (obtained from Section~\ref{subsec:joint_texture_depth_classification}), normalized by the 90th percentile gradient value to ensure robustness against outliers.
\end{enumerate}

We define the BDI $\xi$ as a weighted sum of these normalized factors:

\begin{equation}
    \begin{split}
    \xi = {} & w_1 \cdot D_{phys,norm} + w_2 \cdot O_{disocc,norm} \\
    & + w_3 \cdot D_{max\_disp,norm} + w_4 \cdot T_{comp,norm}
    \end{split}
    \label{eq:BDI}
\end{equation}

The weights reflect each factor's relative importance in causing VSD under large baseline configurations. They can be obtained through grid search on a validation subset of MPEG 3DV sequences.

\subsubsection{BDI-Driven Compensation Mechanism for LS Region Distortion}
\label{subsubsec:BDI_compensation_LS}

To address the limitations of the basic PSD model while maintaining computational efficiency, we propose a BDI-driven compensation mechanism that adapts to large baseline configurations. The BDI serves as an aggregate indicator of synthesis difficulty under large baseline configurations.

Rather than employing a hard threshold approach that would create discontinuous VSDE, we use a smooth transition mechanism based on the following considerations:

\begin{itemize}
    \item Continuity of quality degradation: Rather than degrading abruptly, view synthesis quality degrades gradually with increasing baseline distance. This smooth degradation should be reflected in the VSDE model.
    
    \item Stability of RDO: Discontinuous VSDE functions can lead to unstable bit allocation. A smooth transition ensures that small baseline changes do not cause abrupt shifts in encoding decisions.
\end{itemize}

We employ a sigmoid function for the compensation mechanism due to its favorable properties: S-shaped transition between regimes, bounded output range preventing unrealistic estimates, and computational efficiency with well-defined derivatives.

The proposed compensation mechanism employs multiplicative PSD scaling:

\begin{equation}
    E_{LS,scaled} = E_{LS,PSD} \cdot S(\xi)
    \label{eq:LS_PSD_compensation}
\end{equation}
    
where the scaling function is:

\begin{equation}
    S(\xi) = 1 + \alpha \cdot \sigma(\beta(\xi - \xi_{thresh}))
    \label{eq:BDI_sigmoid}
\end{equation}

\begin{equation}
    \sigma(x) = \frac{1}{1 + e^{-x}}
\end{equation}
    
The parameters can be obtained using a validation subset of MPEG 3DV sequences:
\begin{itemize}
    \item $\xi_{thresh}$: The threshold BDI value that represents the transition point between small and large baseline configurations.
    \item $\alpha$: This parameter determines the maximum multiplicative increment applied to the PSD-based distortion estimation.
    \item $\beta$: A higher $\beta$ creates a sharper transition, while a lower $\beta$ results in a more gradual change.
\end{itemize}

\subsubsection{Final Compensated LS VSDE Model}
\label{subsubsec:final_LS_VSDE}

The total compensated VSDE for LS regions is:

\begin{equation}
    E_{LS,total} = E_{LS,PSD} \cdot (1+\alpha \cdot \sigma(\beta(\xi-\xi_{thresh})))
    \label{eq:final_LS_VSDE}
\end{equation}

where $E_{LS,PSD}$ is obtained by integrating the PSD from equation \eqref{eq:PSD_basic} over all frequencies.

This multiplicative compensation accounts for the collective impact of large baseline effects that violate the PSD model's assumptions. 

This formulation ensures that:

\begin{enumerate}
    \item For small baselines (low BDI), the compensation is minimal, preserving the accuracy of the basic PSD model.
    \item For large baselines (high BDI), the compensation appropriately addresses the increased VSD from geometric effects and disocclusions.
    \item The transition between small and large baselines is smooth and continuous, ensuring stability for RDO.
\end{enumerate}

The computational complexity of this compensation mechanism is minimal, requiring only a few arithmetic operations per frame. Therefore, it preserves the low-complexity advantage of the overall VSDE method while significantly improving accuracy in challenging large baseline configurations.

\subsection{Estimation of VSD Caused by Depth Coding in NS Regions}
\label{subsec:VSDE_NS}

In this section, we study the VSD caused by depth coding in NS regions. These regions violate the WSS assumptions that the frequency-domain PSD-based model relies on. As a result, a different estimation approach is required.

\subsubsection{Property Analysis of NS Regions}

The main challenge in NS regions arises from their lack of statistical homogeneity. LS regions exhibit relatively uniform distortion patterns that can be described by global statistical properties. In contrast, NS regions show highly localized and spatially varying distortion behaviors. Even small depth errors in NS regions can lead to significant misalignment distortions in the synthesized view.

We employ a spatial-domain method for NS region distortion analysis. Our proposed joint texture-depth classification method identifies NS regions more accurately than the texture-only methods \cite{2014TIP_Fang} by capturing important areas around depth discontinuities where the texture is smooth. Although this results in more pixels being classified as NS (approximately 5\% versus 1\% for texture-only methods), the computational complexity of pixel-level processing remains acceptable for real-time applications.

\subsubsection{Taylor Expansion Method}

To address these challenges while maintaining computational efficiency, we propose a Taylor expansion based VSDE method for NS regions. This method reuses gradient computations from Section~\ref{subsec:joint_texture_depth_classification} and adds only second-order derivative calculations, minimizing additional complexity.

Mathematical derivation: Consider the view synthesis process where a pixel at position $(x,y)$ in the virtual view should ideally be warped from texture pixel position $(x',y)$ in the reference view. Due to depth coding errors, the actual warped texture pixel comes from position $(x' + \Delta x, y)$, where $\Delta x$ is the disparity error caused by depth coding.

Let $\hat{T}(x,y)$ represent the pixel value in the reconstructed reference texture image. For a disparity error $\Delta x$ due to depth coding, the ideal and actual texture values can be related through Taylor series expansion:

\begin{equation}
    \begin{split}
        \hat{T}(x' + \Delta x, y) \approx{}& \hat{T}(x',y) + \frac{\partial \hat{T}}{\partial x}\bigg|_{x'}\Delta x \\
        & + \frac{1}{2}\frac{\partial^2 \hat{T}}{\partial x^2}\bigg|_{x'}(\Delta x)^2 + O((\Delta x)^3)
    \end{split}
\end{equation}

where $\frac{\partial \hat{T}}{\partial x}$ is the horizontal gradient of the texture image and $\frac{\partial^2 \hat{T}}{\partial x^2}$ is the second-order horizontal derivative (curvature).

The synthesis error at position $(x,y)$ in the virtual view is the difference between the ideal texture value and the erroneously warped texture value:

\begin{equation}
    \begin{split}
        e(x,y) &= \hat{T}(x',y) - \hat{T}(x' + \Delta x, y) \\
           &\approx -\frac{\partial \hat{T}}{\partial x}\bigg|_{x'}\Delta x - \frac{1}{2}\frac{\partial^2 \hat{T}}{\partial x^2}\bigg|_{x'}(\Delta x)^2
    \end{split}
\end{equation}

For simplicity of notation, we omit the position subscripts:

\begin{equation}
    e(x,y) \approx -\frac{\partial \hat{T}}{\partial x}\Delta x - \frac{1}{2}\frac{\partial^2 \hat{T}}{\partial x^2}(\Delta x)^2
\end{equation}

Computational efficiency through gradients reuse from Section~\ref{subsec:joint_texture_depth_classification}: The first-order derivatives are obtained from the horizontal Sobel operator results computed in Section~\ref{subsec:joint_texture_depth_classification}:

\begin{equation}
    \frac{\partial \hat{T}}{\partial x}(x,y) = G_{x,texture}(x,y)
\end{equation}

where $G_{x,texture}(x,y)$ is the horizontal gradient obtained by applying the horizontal Sobel kernel to the texture image.

The second-order derivatives are computed only for NS pixels using the discrete second derivative operator:

\begin{equation}
    \begin{split}
        \frac{\partial^2 \hat{T}}{\partial x^2}(x,y) ={}& \hat{T}(x+1,y) - 2\hat{T}(x,y) + \hat{T}(x-1,y), \\ 
        &  \quad \forall (x,y) \in NS
    \end{split}
\end{equation}

Derivation of complete distortion estimation: To obtain the estimated VSD, we take the expectation of the squared synthesis error:

\begin{equation}
    E[e^2(x,y)] = E\left[\left(-\frac{\partial \hat{T}}{\partial x}\Delta x - \frac{1}{2}\frac{\partial^2 \hat{T}}{\partial x^2}(\Delta x)^2\right)^2\right]
\end{equation}

Expanding the squared term:

\begin{equation}
\begin{split}
    E[e^2(x,y)] ={}& E\left[\left(\frac{\partial \hat{T}}{\partial x}\right)^2(\Delta x)^2\right] \\
    & + E\left[\frac{1}{4}\left(\frac{\partial^2 \hat{T}}{\partial x^2}\right)^2(\Delta x)^4\right] \\
    & + E\left[\frac{\partial \hat{T}}{\partial x}\frac{\partial^2 \hat{T}}{\partial x^2}(\Delta x)^3\right]
\end{split}
\end{equation}

Assuming that the disparity error $\Delta x$ is independent of the texture gradients and has zero mean (i.e., $E[\Delta x] = 0$), the cross-term vanishes:

\begin{equation}
    E\left[\frac{\partial \hat{T}}{\partial x}\frac{\partial^2 \hat{T}}{\partial x^2}(\Delta x)^3\right] = \frac{\partial \hat{T}}{\partial x}\frac{\partial^2 \hat{T}}{\partial x^2}E[(\Delta x)^3] = 0
\end{equation}

Then we have:

\begin{equation}
    E[e^2(x,y)] = \left(\frac{\partial \hat{T}}{\partial x}\right)^2 E[(\Delta x)^2] + \frac{1}{4}\left(\frac{\partial^2 \hat{T}}{\partial x^2}\right)^2 E[(\Delta x)^4]
\end{equation}

For a zero-mean Laplace distributed random variable, its fourth central moment is equal to six times the square of its variance. The variance of the disparity error is already defined as its second moment, $\nu^2_{\Delta x} = E[(\Delta x)^2]$. Therefore, assuming the disparity errors follow such a distribution, we can express their fourth moment directly in terms of their variance as:

\begin{equation}
    E[(\Delta x)^4] = 6(E[(\Delta x)^2])^2 = 6(\nu_{\Delta x}^2)^2
\end{equation}

The Laplace distribution is a reasonable assumption for depth coding errors as it models sparse, heavy-tailed error distributions commonly observed in video coding \cite{2012TIP_Takahashi}. This distribution captures the characteristic that most pixels have small errors while a few pixels may have larger errors due to coding artifacts.

Substituting $E[(\Delta x)^2] = \nu^2_{\Delta x}$ and using the JEM gradient notation:

\begin{equation}
    \begin{split}
        E[e^2(x,y)] & = \left(\frac{\partial \hat{T}}{\partial x}\right)^2 E[(\Delta x)^2] + \frac{1}{4}\left(\frac{\partial^2 \hat{T}}{\partial x^2}\right)^2 E[(\Delta x)^4] \\
        & = G_{x,texture}^2(x,y) \cdot \nu^2_{\Delta x} + \frac{3}{2}\left(\frac{\partial^2 \hat{T}}{\partial x^2}\right)^2 \cdot (\nu^2_{\Delta x})^2
    \end{split}
\end{equation}

Complete VSDE: The total estimated distortion for NS regions is:

\begin{equation}
    \begin{split}
        E_{NS,Taylor}[Z^2] = {} & \frac{1}{|NS|} \sum_{(x,y) \in NS} \left[ G_{x,texture}^2(x,y) \cdot \nu^2_{\Delta x} \right. \\
        & \left. \quad + \frac{3}{2}\left(\frac{\partial^2 \hat{T}}{\partial x^2}\right)^2 \cdot (\nu^2_{\Delta x})^2 \right]
    \end{split}
\end{equation}

where $\nu^2_{\Delta x} = k^2_i \cdot \nu^2_{\Delta D}$ is the variance of disparity errors, with $k_i$ being the reference view dependent disparity scaling factor in \eqref{eq:horizontal_disparity} and $\nu^2_{\Delta D}$ the variance of depth coding errors.

Large baseline robustness: The Taylor expansion method provides inherent robustness to large baseline configurations through pixel-level accuracy without spatial averaging, automatic scaling via the disparity variance $\nu^2_{\Delta x}$, inclusion of non-linear curvature terms, and limited impact due to small NS region size (NS regions $< 5\%$ of frame generally). Therefore, no additional compensation mechanism is required.

By reusing the gradients from Section~\ref{subsec:joint_texture_depth_classification}, the first-order derivatives provide zero additional cost, making the method both more accurate and computationally efficient for large baseline configurations.

\subsection{Estimation of Blended VSD}
\label{subsec:estimation_blended_VSD}

When synthesizing virtual views using two reference views, the individual VSD estimates from left and right references must be combined to estimate the final blended VSD. Although DIBR theoretically applies linear combination \eqref{eq:linear_combination} for blending pixels in overlapping regions, the actual VSRS implementation creates a more complex synthesis landscape. The resulting VSD cannot be accurately estimated using simple linear combination of individual reference VSDs because: (1) the synthesized view contains distinct region types with different synthesis characteristics, (2) asymmetric baseline configurations create unequal synthesis difficulty between references, and (3) mutual disocclusion regions require hole-filling that introduces additional distortion.

\subsubsection{Theoretical Framework for Multiview VSDE}
\label{subsubsec:multiview_VSDE_theoretical}

In DIBR synthesis with two reference views, the final blended virtual view consists of four distinct region types based on visibility from reference views after warping:

\begin{itemize}
    \item \textbf{Overlapping regions} ($\Omega_{overlap}$): Pixels where both warped intermediate views provide valid data, enabling linear blending
    \item \textbf{Left-only regions} ($\Omega_L$): Pixels where only the left warped view provides valid data, successfully filling disocclusions from the right view
    \item \textbf{Right-only regions} ($\Omega_R$): Pixels where only the right warped view provides valid data, successfully filling disocclusions from the left view
    \item \textbf{Mutual disocclusion regions} ($\Omega_{none}$): Pixels where neither warped view provides valid data, requiring inpainting from surrounding pixels
\end{itemize}

The total VSD is formulated as:

\begin{equation}
    \begin{split}
        E_{dep}[Z^2] 
        &= P_{overlap} \cdot E[e_{overlap}^2] + P_L \cdot E[e_L^2] \\
        &\quad + P_R \cdot E[e_R^2] + P_{none} \cdot E[e_{none}^2]
    \end{split}
    \label{eq:total_vsd_weighted}
\end{equation}

where $P_{overlap}$, $P_L$, $P_R$, and $P_{none}$ are the proportions of each region type (satisfying $P_{overlap} + P_L + P_R + P_{none} = 1$), and $E[e_{i}^2]$ denotes the expected distortion in each region type.

Throughout the blended VSD estimation formulation, $E_{dep}[Z_i^2]$ for $i \in \{L,R\}$ denotes the total depth coding caused VSD when synthesizing the virtual view using reference view $i$ alone, combining contributions from both LS and NS regions as computed in Sections~\ref{sec:VSDE_LS} and~\ref{subsec:VSDE_NS}:

\begin{equation}
    E_{dep}[Z_i^2] = \frac{|LS_i|}{MN} \cdot E_{LS}[Z_i^2] + \frac{|NS_i|}{MN} \cdot E_{NS}[Z_i^2]
    \label{eq:single_ref_total_vsd}
\end{equation}

This quantity is used in:
\begin{itemize}
    \item \textbf{Overlapping regions}: The distortion from each reference is weighted by the blending coefficients $\alpha^2$ and $(1-\alpha)^2$
    \item \textbf{Single-view regions}: The full single-reference distortion applies since only one view contributes
\end{itemize}

This formulation is valid because the LS/NS classification and distortion models apply regardless of whether the reference view is used alone or blended with another view.

\subsubsection{VSDE in Overlapping Regions}

For pixels in overlapping regions, DIBR applies linear blending \eqref{eq:linear_combination}. The distortion in overlapping regions follows:

\begin{equation}
    \begin{split}
    E[e_{overlap}^2] &= \alpha^2 E_{dep}[Z_L^2] + (1-\alpha)^2 E_{dep}[Z_R^2] \\
    &\quad + 2\alpha(1-\alpha)\rho_{LR}\sqrt{E_{dep}[Z_L^2] \cdot E_{dep}[Z_R^2]}
    \end{split}
    \label{eq:overlap_distortion_complete}
\end{equation}

where $\alpha = b_R/b$ is the blending weight from \eqref{eq:blending_weight} and $\rho_{LR}$ is the correlation coefficient between the depth coding caused VSDs from left and right reference views. For independently encoded reference views, the depth coding errors are largely uncorrelated, allowing us to assume $\rho_{LR} \approx 0$. This assumption is reasonable because:

\begin{itemize}
    \item The left and right views are encoded independently without inter-view prediction
    \item Quantization errors in each view behave as independent zero-mean noise
    \item Any structural correlations are typically weak compared to the independent noise components
\end{itemize}

Therefore, \eqref{eq:overlap_distortion_complete} simplifies to:

\begin{equation}
    E[e_{overlap}^2] \approx \alpha^2 E_{dep}[Z_L^2] + (1-\alpha)^2 E_{dep}[Z_R^2]
    \label{eq:overlap_distortion}
\end{equation}

\subsubsection{VSDE in Single-View Regions}

Single-view regions ($\Omega_L$ and $\Omega_R$) are areas where one reference view successfully provides pixel data that fills disocclusions from the other view. By definition, these regions contain no final holes - all pixels are successfully warped from the available reference view. The distortion in these regions is purely the depth coding distortion from warping:

\begin{equation}
    E[e_L^2] = E_{dep}[Z_L^2], \quad E[e_R^2] = E_{dep}[Z_R^2]
    \label{eq:single_view_distortion}
\end{equation}

where $E_{dep}[Z_L^2]$ and $E_{dep}[Z_R^2]$ are the single-view VSDE from \eqref{eq:single_ref_total_vsd}.

\subsubsection{VSDE in Mutual Disocclusion Regions}

Mutual disocclusion regions ($\Omega_{none}$) occur when both warped intermediate views have holes at the same pixel location, particularly significant in large baseline configurations. For these regions, DIBR employs hole-filling algorithms. We implement a variance-based hole-filling approach in our modified View Synthesis Reference Software (VSRS) \cite{2013MPEG_Wegner}, ensuring consistency between VSDE and synthesis.

Following this variance-based approach for computational efficiency in frame-level estimation, the distortion is modeled as:

\begin{equation}
    E[e_{none}^2] = \frac{\nu^2_{local,L} + \nu^2_{local,R}}{2}
    \label{eq:mutual_disocclusion_distortion}
\end{equation}

where $\nu^2_{local,i}$ is the local variance of texture pixels adjacent to depth discontinuities in reference view $i$, representing the difficulty of inpainting in regions with high texture variation.

\subsubsection{Region Proportion Estimation}

We estimate region proportions using geometric relationships derived from camera configuration and depth discontinuities.

\textbf{Single-view region proportions:}

The disocclusion areas $O_{disocc,L}$ and $O_{disocc,R}$ are already computed in the BDI framework (Section~\ref{subsubsec:BDI}, equation~\eqref{eq:disocclusion_factor}):

\begin{equation}
    O_{disocc,i} = \frac{1}{WH}\sum_{e \in \mathcal{E}_i} W_{disocc,i}(e) \cdot h_e, \quad i \in \{L,R\}
    \label{eq:reuse_disocc_from_bdi}
\end{equation}

where $W_{disocc,i}(e) = f \cdot b_i |1/z_{fg,e} - 1/z_{bg,e}|$ is the disocclusion width at edge $e$. From geometric principles, single-view regions occur where one reference successfully fills the other's disocclusions:

\begin{equation}
\begin{split}
    P_L &= O_{disocc,R} \\
    P_R &= O_{disocc,L}
\end{split}
\label{eq:single_view_proportions}
\end{equation}

This computational reuse maintains the low-complexity advantage of the overall VSDE method.

\textbf{Mutual disocclusion region proportion:}

Mutual disocclusions occur where disocclusion regions from both views spatially overlap. For 1D parallel camera configurations, we match depth edges between reference views by projecting to virtual view coordinates. Edges are matched if:
\begin{equation}
    |x_V(e_L) - x_V(e_R)| < \tau_{spatial}, \quad |y(e_L) - y(e_R)| < \tau_{vertical}
\end{equation}
where $\tau_{spatial} = 2-3$ pixels accounts for depth quantization effects and $\tau_{vertical} = 1$ pixel enforces vertical alignment from the 1D parallel geometry. These geometry-derived thresholds require no training or tuning.

For matched edges creating opposing disocclusions:

\begin{equation}
    W_{overlap}(e) = \begin{cases}
        \min(W_{disocc,L}(e), W_{disocc,R}(e)) & \text{if opposing} \\
        & \text{gradients} \\
        0 & \text{otherwise}
    \end{cases}
    \label{eq:overlap_width}
\end{equation}

where edges have "opposing gradients" when $\text{sgn}(\nabla_x D_L(e)) \neq \text{sgn}(\nabla_x D_R(e))$.

The mutual disocclusion proportion is:
\begin{equation}
    P_{none} = \frac{1}{WH}\sum_{e \in \mathcal{E}_L \cap \mathcal{E}_R} W_{overlap}(e) \cdot h_e
    \label{eq:mutual_disocc_proportion}
\end{equation}

Note that $P_L$ and $P_R$ sum all disocclusions from each view independently, while $P_{none}$ sums only matched, overlapping disocclusions from edges visible in both views.

\textbf{Overlapping region proportion:}

By conservation property:
\begin{equation}
    P_{overlap} = 1 - P_L - P_R - P_{none}
    \label{eq:overlap_proportion}
\end{equation}

All computations reuse previously calculated quantities, maintaining the low-complexity advantage of the overall VSDE method. The computational cost of overlap calculation is $O(|\mathcal{E}|)$ where $|\mathcal{E}|$ is the number of depth edges, typically $\ll WH$ (less than 1-5\% of pixels).

\subsubsection{Complete Blended VSD Formulation}

Combining all components, the total depth coding caused VSD is:

\begin{equation}
    \begin{split}
        E_{dep}[Z^2] &= P_{overlap} \cdot [\alpha^2 E_{dep}[Z_L^2] + (1-\alpha)^2 E_{dep}[Z_R^2]] \\
        &\quad + P_L \cdot E_{dep}[Z_L^2] + P_R \cdot E_{dep}[Z_R^2] \\
        &\quad + P_{none} \cdot \frac{\nu^2_{local,L} + \nu^2_{local,R}}{2}
    \end{split}
    \label{eq:complete_blended_vsd_final}
\end{equation}

where all region proportions are computed geometrically without empirical parameters, ensuring the method generalizes across different sequences and camera configurations.

\subsection{Overall VSDE}
\label{subsec:overall_VSDE}

The complete VSDE method integrates VSDs estimated from texture coding, LS regions, and NS regions to provide a comprehensive framework for VSDE. This section presents the overall mathematical formulation that combines these modules.

\subsubsection{Overall VSDE Framework}

The total estimated VSD follows the decomposition principle established in Section~\ref{subsec:analysis_coding_error_impact}:

\begin{equation}
    E[V^2] \approx E_{tex}[N^2] + E_{dep}[Z^2]
\end{equation}

where $E[V^2]$ represents the total expected VSD in MSE metric, $E_{tex}[N^2]$ is the expected texture coding caused VSD from \eqref{eq:texture_VSD_simplified}, and $E_{dep}[Z^2]$ is the expected depth coding caused VSD from \eqref{eq:complete_blended_vsd_final}.

\subsubsection{VSDE Integration for Single Reference View}

For a single reference view $i \in \{L, R\}$, the total depth coding caused VSDE is the weighted sum of LS and NS regional VSDE contributions:

\begin{equation}
    E_{dep}[Z^2_i] = \frac{|LS_i|}{MN} \cdot E_{LS}[Z^2_i] + \frac{|NS_i|}{MN} \cdot E_{NS}[Z^2_i]
\end{equation}

where $|LS_i|$ and $|NS_i|$ are the numbers of pixels classified as LS and NS regions respectively in reference view $i$, $MN$ is the total frame size, and the weighting factors represent the fractional areas of each region type.

The LS regional distortion $E_{LS}[Z^2_i]$ is estimated using the compensated PSD-based model from Section~\ref{sec:VSDE_LS}:

\begin{equation}
    E_{LS}[Z^2_i] = E_{LS,PSD} \cdot S(\xi_i)
\end{equation}

where $E_{LS,PSD}$ is the basic PSD model from \eqref{eq:PSD_basic}, and $S(\xi_i) = 1 + \alpha \cdot \sigma(\beta(\xi_i - \xi_{thresh}))$ is the sigmoid compensation function from \eqref{eq:BDI_sigmoid}.

The NS region distortion $E_{NS}[Z^2_i]$ is estimated using the Taylor expansion method from Section~\ref{subsec:VSDE_NS}:

\begin{equation}
    \begin{split}
        E_{NS}[Z^2_i] = {} & \frac{1}{|NS_i|} \sum_{(x,y) \in NS_i} \left[ G_{x,texture}^2(x,y) \cdot \nu^2_{\Delta x} \right. \\
        & \left. \quad + \frac{3}{2}\left(\frac{\partial^2 \hat{T}}{\partial x^2}(x,y)\right)^2 \cdot (\nu^2_{\Delta x})^2 \right]
    \end{split}
\end{equation}

where $G_{x,texture}^2(x,y)$ is the squared horizontal gradient from Section~\ref{subsec:joint_texture_depth_classification}, $\nu^2_{\Delta x} = k^2_i \cdot \nu^2_{\Delta D}$ is the disparity error variance, and the factor $\frac{3}{2}$ accounts for the fourth moment of Laplace distributed disparity errors.

\subsubsection{VSDE of Multiview Combination}

When using two reference views, the depth coding caused VSD is estimated using the region-based formulation from Section~\ref{subsec:estimation_blended_VSD}:

\begin{equation}
    \begin{split}
        E_{dep}[Z^2] &= P_{overlap} \cdot [\alpha^2 E_{dep}[Z_L^2] + (1-\alpha)^2 E_{dep}[Z_R^2]] \\
        &\quad + P_L \cdot E_{dep}[Z_L^2] + P_R \cdot E_{dep}[Z_R^2] \\
        &\quad + P_{none} \cdot \frac{\nu^2_{local,L} + \nu^2_{local,R}}{2}
    \end{split}
    \label{eq:multiview_vsd}
\end{equation}

where:
\begin{itemize}
    \item $\alpha = b_R/b$ is the blending coefficient from \eqref{eq:blending_weight}
    \item $P_{overlap}, P_L, P_R, P_{none}$ are region proportions computed geometrically:
    \begin{itemize}
        \item $P_L = O_{disocc,R}$, $P_R = O_{disocc,L}$ from BDI computation
        \item $P_{none}$ from geometric overlap of matched edges
        \item $P_{overlap} = 1 - P_L - P_R - P_{none}$
    \end{itemize}
    \item $E_{dep}[Z_L^2]$, $E_{dep}[Z_R^2]$ are single-view VSDE including both LS and NS contributions from \eqref{eq:single_ref_total_vsd}
    \item $\nu^2_{local,i}$ are local texture variances near depth discontinuities, representing variance-based hole-filling distortion
\end{itemize}

This region-based formulation accounts for the heterogeneous nature of blended view synthesis, where different pixel locations experience different synthesis conditions based on reference view visibility.

\subsubsection{Complete VSDE Formulation}

The complete VSDE method integrates VSDs estimated from texture coding, LS regions, and NS regions to provide a comprehensive framework for VSDE. The total estimated VSD follows the decomposition principle established in Section~\ref{subsec:analysis_coding_error_impact}:

\begin{equation}
    E[V^2] \approx E_{tex}[N^2] + E_{dep}[Z^2]
    \label{eq:total_vsd_decomposition}
\end{equation}

where $E[V^2]$ represents the total expected VSD in MSE metric, $E_{tex}[N^2]$ is the texture coding caused VSD from \eqref{eq:texture_VSD_simplified}, and $E_{dep}[Z^2]$ is the depth coding caused VSD from \eqref{eq:multiview_vsd}.

Substituting all components, the complete VSDE becomes:

\begin{equation}
    \begin{split}
        E[V^2] = {} & \alpha^2 E[(T_L - \hat{T}_L)^2] + (1-\alpha)^2 E[(T_R - \hat{T}_R)^2] \\
        & + P_{overlap} \cdot [\alpha^2 E_{dep}[Z^2_L] + (1-\alpha)^2 E_{dep}[Z^2_R]] \\
        & + P_L \cdot E_{dep}[Z^2_L] + P_R \cdot E_{dep}[Z^2_R] \\
        & + P_{none} \cdot \frac{\nu^2_{local,L} + \nu^2_{local,R}}{2}
    \end{split}
    \label{eq:complete_vsd_final}
\end{equation}

where for each reference view $i \in \{L,R\}$:
\begin{equation}
    E_{dep}[Z^2_i] = \frac{|LS_i|}{MN} \cdot E_{LS}[Z^2_i] + \frac{|NS_i|}{MN} \cdot E_{NS}[Z^2_i]
    \label{eq:depth_vsd_components}
\end{equation}

The LS regional distortion $E_{LS}[Z^2_i]$ uses the BDI-compensated PSD model from Section~\ref{sec:VSDE_LS}:
\begin{equation}
    E_{LS}[Z^2_i] = E_{LS,PSD} \cdot (1 + \alpha \cdot \sigma(\beta(\xi_i - \xi_{thresh})))
    \label{eq:ls_with_bdi}
\end{equation}

The NS region distortion $E_{NS}[Z^2_i]$ uses the Taylor expansion model from Section~\ref{subsec:VSDE_NS}:
\begin{equation}
    \begin{split}
        E_{NS}[Z^2_i] = {} & \frac{1}{|NS_i|} \sum_{(x,y) \in NS_i} \left[ G_{x,texture}^2(x,y) \cdot \nu^2_{\Delta x} \right. \\
        & \left. \quad + \frac{3}{2}\left(\frac{\partial^2 \hat{T}}{\partial x^2}(x,y)\right)^2 \cdot (\nu^2_{\Delta x})^2 \right]
    \end{split}
    \label{eq:ns_taylor}
\end{equation}

The formulation maintains computational efficiency by:
\begin{itemize}
    \item Reusing edge detection and disocclusion computations from BDI
    \item Reusing gradient computations from LS/NS classification
    \item Computing region proportions geometrically without iteration
    \item Avoiding actual DIBR synthesis for estimation
\end{itemize}

\subsubsection{Computational Complexity Analysis}

The proposed VSDE method maintains computational efficiency suitable for real-time applications through strategic algorithm design and computational reuse. The overall complexity is $O(MN)$ per frame, where $M \times N$ is the frame resolution.

\textbf{Component-wise complexity:}
\begin{itemize}
    \item Joint texture-depth classification: $O(MN)$ for Sobel operators and Otsu thresholding
    \item BDI computation: $O(MN)$ for edge detection, $O(|\mathcal{E}|)$ for disocclusion summation where $|\mathcal{E}| \ll MN$ (typically 1-5\% of pixels)
    \item LS region PSD analysis: $O(|LS|) \approx O(0.95MN)$ for frequency-domain integration
    \item NS region Taylor expansion: $O(|NS|) \approx O(0.05MN)$ for spatial-domain computation
    \item Region proportion estimation: $O(|\mathcal{E}|)$ for edge matching and geometric overlap calculation
    \item Multiview blending combination: $O(1)$ for weighted summation
\end{itemize}

\textbf{Computational reuse strategy:}
\begin{itemize}
    \item Gradient computations from classification are reused in NS analysis and mutual disocclusion detection
    \item Disocclusion areas from BDI ($O_{disocc,L}$, $O_{disocc,R}$) are reused for region proportions ($P_L$, $P_R$)
    \item Edge detection results are reused across BDI, classification, and proportion estimation
    \item Local variance calculations serve both BDI and hole-filling distortion estimation
\end{itemize}

\textbf{Efficiency through DIBR avoidance:} By directly estimating VSD without performing actual view synthesis, the method avoids computationally expensive operations including geometric transformations ($O(MN)$ per view), warping competition resolution, iterative hole-filling algorithms, and pixel-level blending. This represents a significant improvement over actual DIBR synthesis which requires multiple $O(MN)$ passes with complex operations.

The linear scalability ensures the method remains practical for high-resolution content and real-time applications, with total computational overhead dominated by texture and depth map processing rather than synthesis operations.

\subsubsection{Implementation Considerations}

The modular design enables several optimization strategies:

\begin{itemize}
    \item Parallel processing: LS and NS region analyses can be performed simultaneously, and left/right reference view processing can be parallelized.
    
    \item Adaptive precision: For extreme real-time requirements, the NS region Taylor expansion can be limited to first-order terms, trading minimal accuracy for additional speed.
    
    \item Dynamic configuration: The training-free nature allows dynamic adaptation to changing camera configurations without preprocessing, making it suitable for adaptive streaming applications.
\end{itemize}

Therefore, the proposed VSDE method provides an effective balance between computational efficiency and estimation accuracy, making it practical for real-time 3DV applications such as interactive FVV systems and adaptive 3D streaming services, particularly in challenging large baseline configurations.

\section{Experiments}
\label{sec:experiments}

This section presents a comprehensive experimental validation of the proposed VSDE method. The experiments are designed to systematically evaluate individual modules, overall system performance, robustness across different baseline configurations, and asymmetric QP setups. The validation framework demonstrates the effectiveness of each novel contribution while comparing against existing methods under various challenging conditions. And we particularly focus on large baseline configurations that represent the core contribution of this work.

\subsection{Experimental Setup}
\label{subsec:experimental_setup}

We validate our method using the standard MPEG 3DV test sequences \textit{PoznanHall2}, \textit{PoznanStreet}, \textit{CarPark} \cite{2020MPEG_Mieloch}, \textit{Kendo}, \textit{Balloons}, \textit{BookArrival}, \textit{GTFly}, \textit{UndoDancers}. These sequences provide different scene characteristics including varying texture complexities, depth range distributions, and object motion patterns, which can thoroughly test the robustness of the proposed VSDE method across different content types. To ensure unbiased evaluation, we use a validation subset approach: the sequences \textit{CarPark}, \textit{UndoDancers}, \textit{GTFly}, and \textit{BookArrival} are used as a validation set for obtaining both the BDI weights (Section~\ref{subsubsec:BDI}) and the sigmoid compensation parameters (Section~\ref{subsubsec:BDI_compensation_LS}). Specifically, through grid search on this validation subset, we obtained: BDI weights $w_1=0.3$, $w_2=0.4$, $w_3=0.2$, $w_4=0.1$, and sigmoid parameters $\xi_{thresh}=0.4$, $\alpha=1.5$, $\beta=10$. The remaining sequences \textit{PoznanHall2}, \textit{PoznanStreet}, \textit{Kendo}, and \textit{Balloons} are reserved exclusively for performance evaluation, ensuring that our results are not influenced by parameter optimization.

We follow the standard MPEG 3DV 2-view test case configurations \cite{2011MPEG_N12036}, where two reference views are used to synthesize virtual views between them. The ground truth for MSE calculation is obtained from virtual views synthesized using uncompressed texture and depth sequences, consistent with established MPEG 3DV testing practices. This ensures fair comparison with existing methods.

Encoding configuration: Both texture and depth sequences are encoded using JMVC 8.3.1 reference software. Each Group-of-Pictures (GOP) consists of an anchor frame followed by hierarchically coded B frames. Inter-view prediction and rate-control are disabled to ensure precise QP control and maintain consistent experimental conditions across all test cases. QP values for texture ($QP_T$) and depth ($QP_D$) range from 28 to 48, providing a comprehensive range of coding qualities from high to low bitrates.

View synthesis configuration: Virtual views are generated using the VSRS 4.0 with 1D parallel camera configuration \cite{2013MPEG_Wegner}. Linear blending is employed for combining left and right intermediate views, with a local variance-based approach implemented for disocclusion/hole filling (as described in Section~\ref{subsec:estimation_blended_VSD}). This ensures consistency between our hole-filling distortion estimation and the actual view synthesis process.

Baseline configurations: To thoroughly evaluate large baseline configurations performance, we test multiple baseline distance configurations:
\begin{itemize}
    \item Small baseline: Baseline distances under 2.0 camera units, representing typical dense camera configurations
    \item Medium baseline: Baseline distances from 2.0 to 4.0 camera units, representing moderate sparse camera configurations
    \item Large baseline: Baseline distances from 4.0 to 8.0 camera units, representing challenging sparse camera configurations common in bandwidth-constrained applications
\end{itemize}

Performance metrics: The primary performance metrics are MSE between estimated and actual VSD values, since our method directly predicts VSD in MSE metric. We also employ:

\begin{itemize}
    \item Pearson Correlation Coefficient (PCC): Measures the linear correlation between estimated and actual VSD values:
    \begin{equation}
        \text{PCC} = \frac{\sum_{i=1}^{N}(y_i - \bar{y})(\hat{y}_i - \bar{\hat{y}})}{\sqrt{\sum_{i=1}^{N}(y_i - \bar{y})^2}\sqrt{\sum_{i=1}^{N}(\hat{y}_i - \bar{\hat{y}})^2}}
    \end{equation}
    where $y_i$ is the actual VSD, $\hat{y}_i$ is the estimated VSD, and $N$ is the number of test cases.
    
    \item Root Mean Squared Error (RMSE): Quantifies the estimation accuracy in MSE units:
    \begin{equation}
        \text{RMSE} = \sqrt{\frac{1}{N}\sum_{i=1}^{N}(y_i - \hat{y}_i)^2}
    \end{equation}
\end{itemize}

The overall performance of the proposed method is validated by comparing its VSDE against the ground truth MSE. We also compare it against two ablations of our own method: one without the BDI-driven compensation mechanism and using simple linear blending estimation instead of our proposed region-based blending estimation strategy. 

In the subsequent sections, we present a comprehensive evaluation of the proposed method. We first conduct ablation studies to validate the effectiveness of each key module, then we evaluate its overall performance. Furthermore, we conduct a comparative analysis against two existing similar models: Yuan's polynomial model \cite{2014TCSVT_Yuan} and Fang's analytical model \cite{2014TIP_Fang}. We compare against Yuan's and Fang's methods as they represent the most recent frame-level VSDE approaches that directly estimate MSE, making them the most comparable to our method. More recent works either focus on pixel-level processing \cite{2022TMM_Jin}, subjective quality assessment \cite{2019TIP_Dragana, 2020TIP_Wang}, or require extensive pre-training \cite{2024ASC_Shi}.

\subsection{Individual Module Validation}

In this section, we design isolated experiments to validate the effectiveness of the three modules of our proposed VSDE method respectively: the joint texture-depth classification method, the BDI design, and the region-based blending estimation strategy.

\subsubsection{Joint Texture-Depth Classification Effectiveness}
\label{subsubsec:experimental_joint_texture-depth}

This experiment validates our joint texture-depth classification method against the texture-only gradient-based approach from \cite{2014TIP_Fang}.

We test on \textit{Kendo} and \textit{Balloons} sequences with small to medium baseline configurations (V1+V3$\rightarrow$V2, V1+V5$\rightarrow$V2, V1+V5$\rightarrow$V3) where both methods' assumptions remain valid. To isolate classification performance, we:
\begin{enumerate}
    \item Generate LS/NS masks using both classification methods
    \item Apply the same VSDE model from \cite{2014TIP_Fang} to both masks
    \item Compare estimated VSD against ground truth
    \item Use uncompressed texture images to ensure differences arise solely from classification accuracy
\end{enumerate}

\begin{figure}[t!]
    \centering
    \setlength{\tabcolsep}{1pt}
    \begin{tabular}{cc}
        \includegraphics[width=0.48\columnwidth]{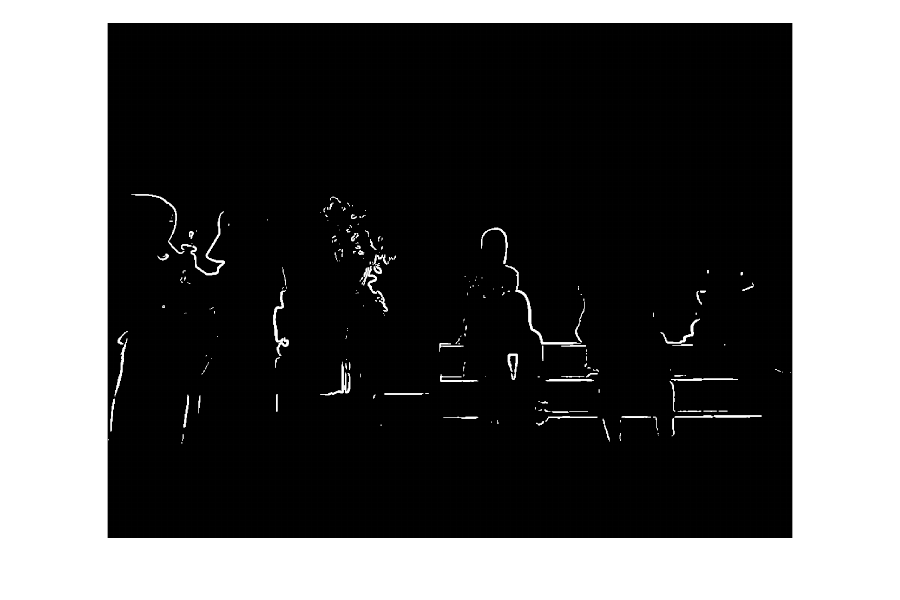} &
        \includegraphics[width=0.48\columnwidth]{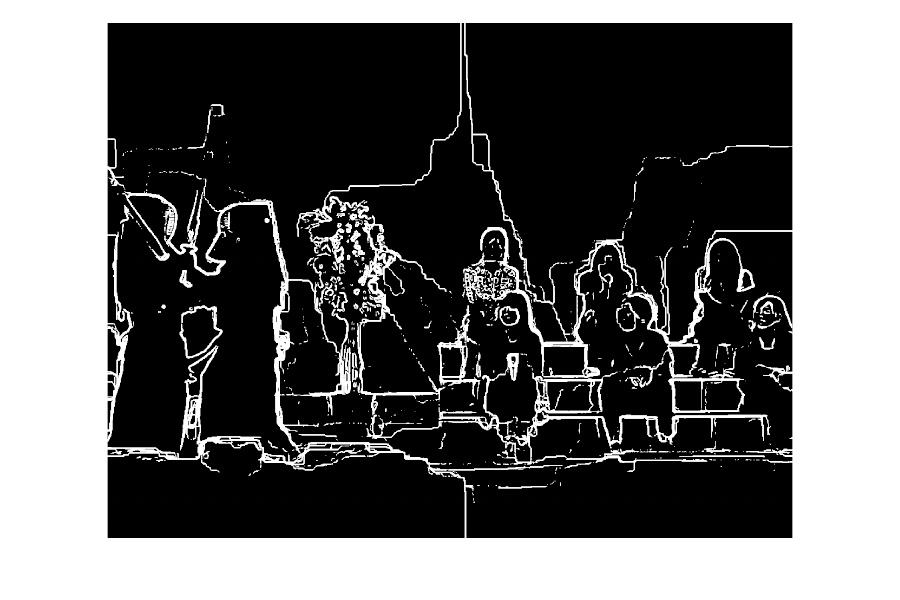} \\
        \multicolumn{2}{c}{{(a) Kendo View 3, Frame 43}} \\[3pt]
        \includegraphics[width=0.48\columnwidth]{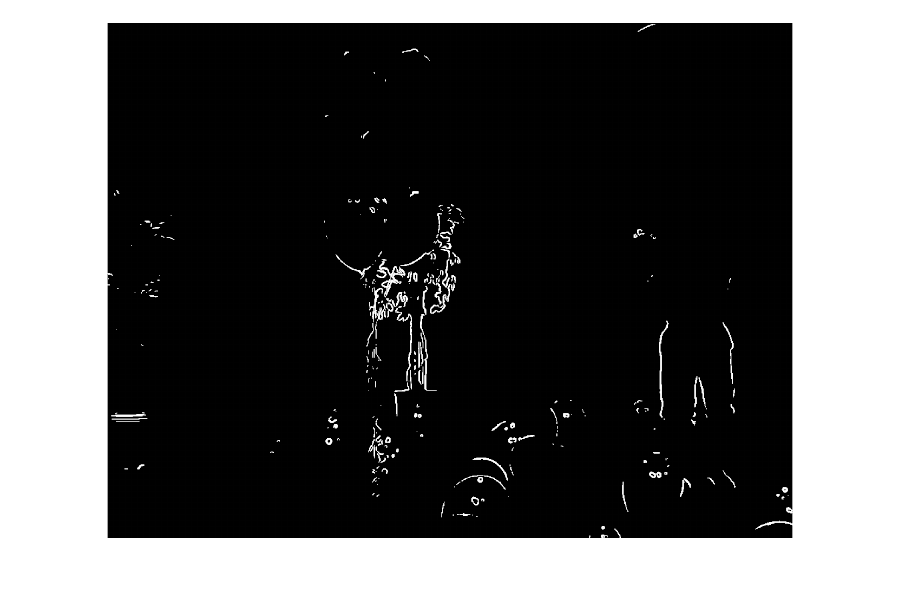} &
        \includegraphics[width=0.48\columnwidth]{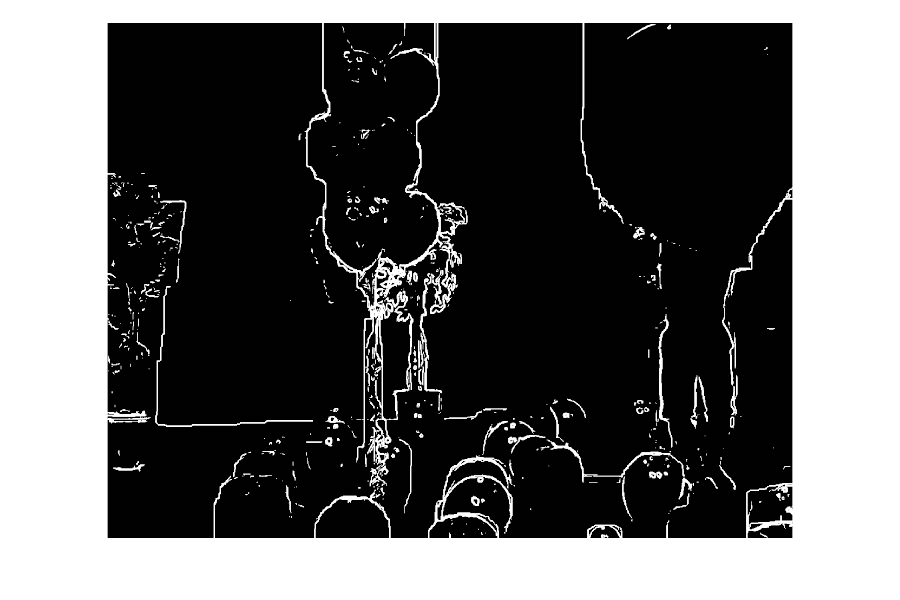} \\
        \multicolumn{2}{c}{{(b) Balloons View 3, Frame 43}}
    \end{tabular}
    \caption{LS/NS classification masks comparison. Left column: Texture-only \cite{2014TIP_Fang} method. Right column: Joint texture-depth classification (proposed method). Both sequences shown at View 3, Frame 43.}
    \label{fig:masks_comparison}
\end{figure}

\textbf{Visual analysis:} Fig.~\ref{fig:masks_comparison} reveals substantial differences between methods. The texture-only method \cite{2014TIP_Fang} identifies only 1.1\% of pixels as NS regions on average, missing critical depth discontinuities in smooth texture areas. Our joint approach identifies 5.0\% as NS regions, correctly capturing both texture edges and depth boundaries—a 4× increase that better represents actual synthesis challenges.

\textbf{Quantitative performance:} Table~\ref{tab:joint_texture-depth_classification_results} shows our method achieves 62.1\% RMSE reduction (0.59 to 0.22) averaged across depth QP 28-48. The improvement is most pronounced in medium baseline cases (V1+V5), where accurate NS region identification becomes critical for VSDE accuracy.

\begin{figure}[t!]
    \centering
    \setlength{\tabcolsep}{1pt}
    \begin{tabular}{cc}
        \includegraphics[width=0.48\columnwidth]{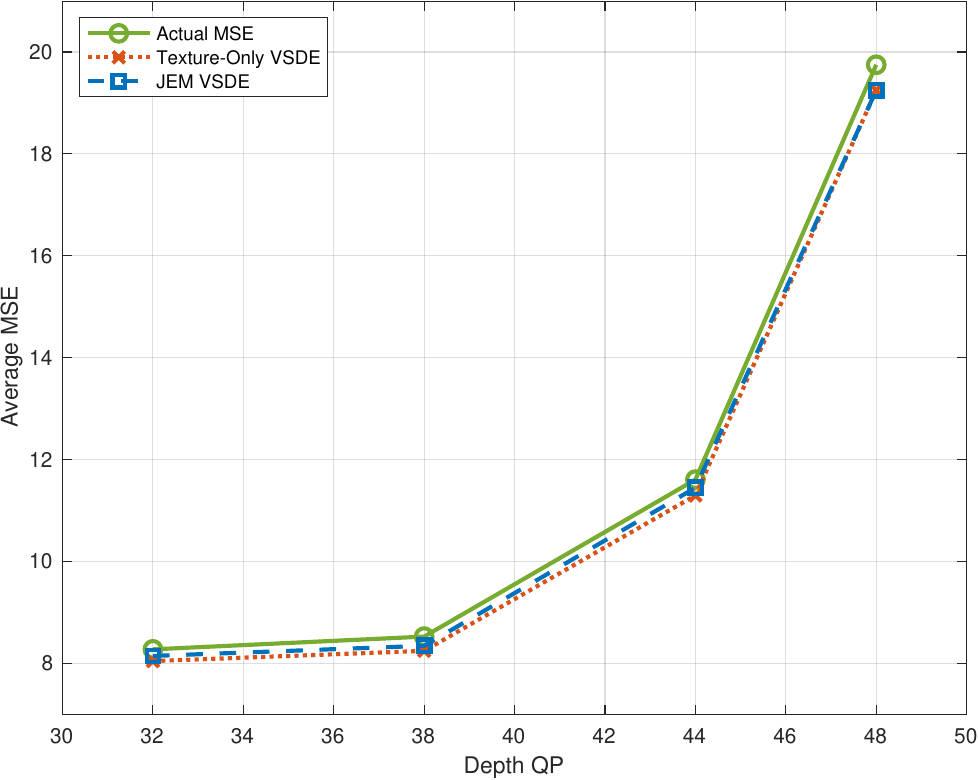} &
        \includegraphics[width=0.48\columnwidth]{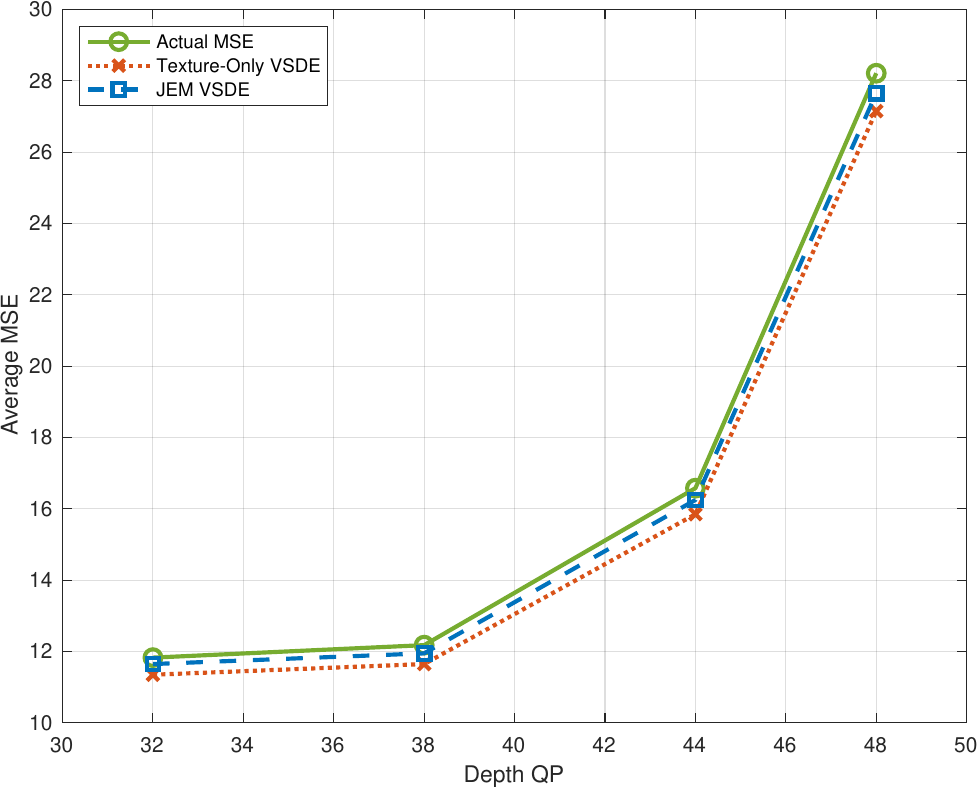} \\
        {(a) V1+V3$\rightarrow$V2} & 
        {(b) V1+V5$\rightarrow$V2} \\[3pt]
        \includegraphics[width=0.48\columnwidth]{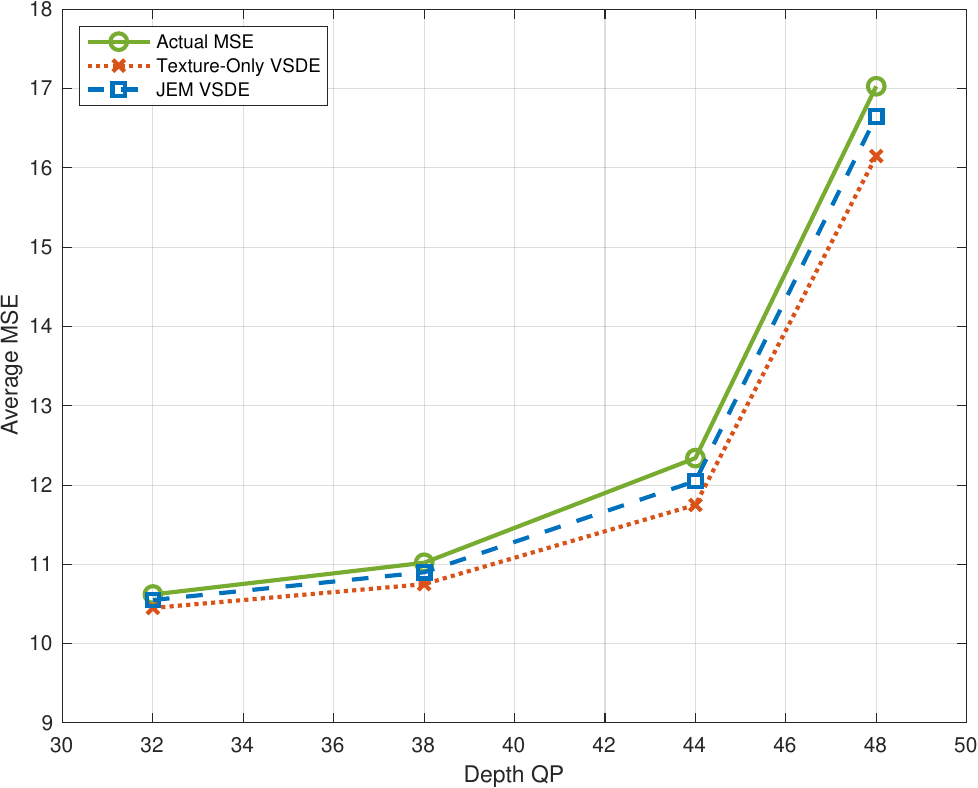} &
        \includegraphics[width=0.48\columnwidth]{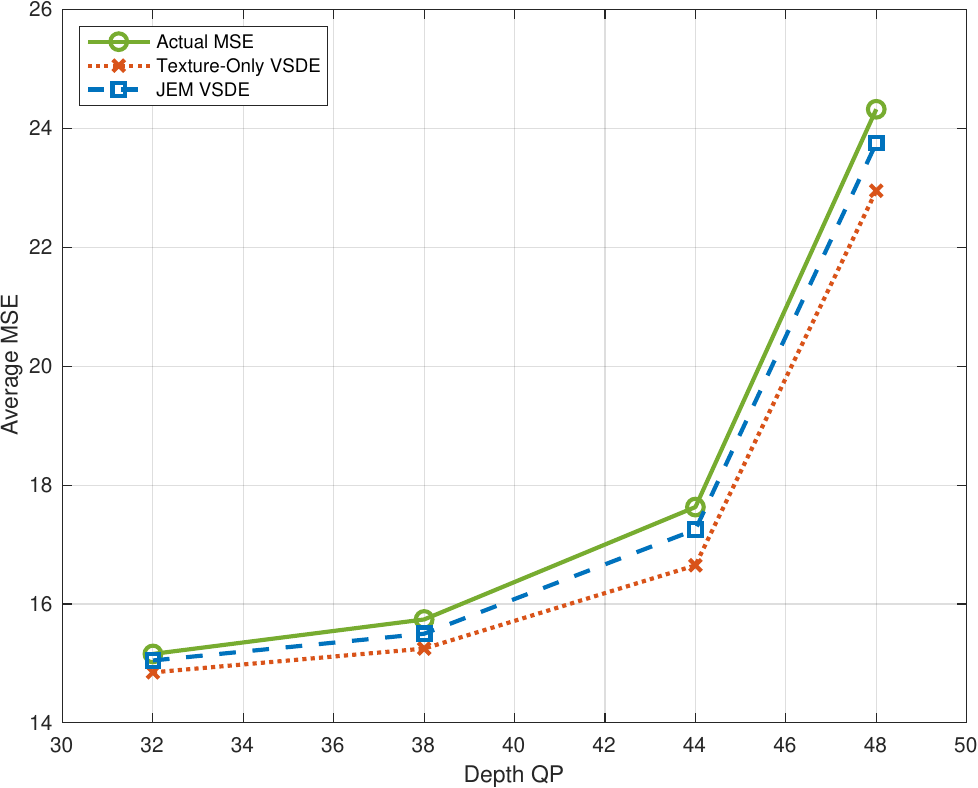} \\
        {(c) V1+V3$\rightarrow$V2} & 
        {(d) V1+V5$\rightarrow$V2}
    \end{tabular}
    \caption{VSDE performance with LS/NS separation. Top row: Kendo sequence. Bottom row: Balloons sequence.}
    \label{fig:vsde_comparison}
\end{figure}

\textbf{VSDE performance curves:} Fig.~\ref{fig:vsde_comparison} demonstrates consistent improvements across all depth QP values, with larger gains at high QP where compression artifacts amplify classification importance. The performance hierarchy "texture-only < joint classification < complete proposed method" validates that improved classification directly enhances VSDE accuracy.

While improvements are moderate in these small-to-medium baseline tests, we expect more substantial gains in large baseline configurations where depth discontinuities have greater synthesis impact.

\begin{table}[t!]
\centering
\renewcommand\arraystretch{1.3}
\caption{Joint Texture-Depth Classification Performance}
\label{tab:joint_texture-depth_classification_results}
\resizebox{\columnwidth}{!}{%
\begin{tabular}{llcccc}
\hline
\textbf{Sequence} & \textbf{Configuration} & \textbf{Method} & \textbf{RMSE} & \textbf{LS\%} & \textbf{NS\%} \\
\hline
\hline
\multirow{6}{*}{Kendo} 
& V1+V3$\rightarrow$V2 & Texture$^a$ & 0.47 & 99.2 & 0.8 \\
& & Proposed & 0.17 & 93.1 & 6.9 \\
\cline{2-6}
& V1+V5$\rightarrow$V2 & Texture & 0.66 & 98.9 & 1.1 \\
& & Proposed & 0.23 & 94.7 & 5.3 \\
\cline{2-6}
& V1+V5$\rightarrow$V3 & Texture & 0.68 & 99.0 & 1.0 \\
& & Proposed & 0.25 & 93.8 & 6.2 \\
\hline
\hline
\multirow{6}{*}{Balloons} 
& V1+V3$\rightarrow$V2 & Texture & 0.45 & 99.1 & 0.9 \\
& & Proposed & 0.17 & 96.2 & 3.8 \\
\cline{2-6}
& V1+V5$\rightarrow$V2 & Texture & 0.63 & 98.8 & 1.2 \\
& & Proposed & 0.25 & 95.8 & 4.2 \\
\cline{2-6}
& V1+V5$\rightarrow$V3 & Texture & 0.65 & 98.6 & 1.4 \\
& & Proposed & 0.27 & 96.5 & 3.5 \\
\hline
\hline
\multicolumn{2}{c}{\textbf{Average}} & Texture & \textbf{0.59} & \textbf{98.9} & \textbf{1.1} \\
\multicolumn{2}{c}{} & Proposed & \textbf{0.22} & \textbf{95.0} & \textbf{5.0} \\
\multicolumn{2}{c}{\textbf{Improvement}} & & \textbf{-62.1\%} & \textbf{--} & \textbf{--} \\
\hline
\multicolumn{6}{l}{$^a$Texture: Texture-Only method from \cite{2014TIP_Fang}}\\
\end{tabular}
}
\end{table}

\subsubsection{BDI Validation}
\label{subsubsec:experimental_BDI}

This experiment validates our proposed BDI by examining whether its individual factors correlate with the basic PSD model's estimation errors.

We use large baseline configuration (V0+V8$\rightarrow$V4 with 8.0 camera units) where the basic PSD model's limitations are most apparent. For each frame in \textit{PoznanHall2} and \textit{PoznanStreet} sequences, we:
\begin{enumerate}
    \item Compute the actual VSD from ground truth
    \item Estimate VSD using the basic PSD model (equation~\eqref{eq:PSD_basic}) without compensation
    \item Calculate the absolute estimation error: $e = |\text{VSD}_{\text{actual}} - \text{VSD}_{\text{PSD}}|$
    \item Extract the four normalized BDI factors for that frame
\end{enumerate}

We then compute the Pearson correlation coefficient between each factor's values and the estimation errors across all frames. A higher correlation indicates that factor better predicts when the basic PSD model needs compensation.

\textbf{Individual factor correlation analysis:} Table~\ref{tab:bdi_individual_factors} shows that disocclusion area achieves the highest correlation (0.89-0.91), validating its assignment of the highest weight ($w_2=0.4$). Physical baseline distance (0.82-0.84) and maximum disparity (0.73-0.75) show moderate correlations, while texture complexity (0.55-0.61) has the weakest correlation. The combined BDI achieves 0.92-0.93 correlation, confirming that the weighted combination outperforms any individual factor.

\textbf{Weight configuration validation:} Table~\ref{tab:bdi_weight_analysis} compares our proposed weights against three alternatives. Our configuration achieves the best performance (PCC=0.92), while equal weighting performs worst (PCC=0.85), demonstrating the importance of appropriate factor prioritization.

\begin{table}[t!]
\centering
\renewcommand\arraystretch{1.3}
\caption{Individual BDI Factor Correlations with Uncompensated Estimation Error}
\label{tab:bdi_individual_factors}
\resizebox{\columnwidth}{!}{%
\begin{tabular}{lccc}
\hline
\textbf{BDI Factor} & \textbf{PoznanHall2} & \textbf{PoznanStreet} & \textbf{Weight} \\
& \textbf{PCC} & \textbf{PCC} & \textbf{($w_i$)} \\
\hline
\hline
Physical Baseline ($D_{phys,norm}$) & 0.84 & 0.82 & 0.3 \\
Disocclusion Area ($O_{disocc,norm}$) & 0.89 & 0.91 & 0.4 \\
Maximum Disparity ($D_{maxdisp,norm}$) & 0.73 & 0.75 & 0.2 \\ 
Texture Complexity ($T_{comp,norm}$) & 0.55 & 0.61 & 0.1 \\
\hline
\textbf{Combined BDI} & \textbf{0.92} & \textbf{0.93} & \textbf{--} \\
\hline
\multicolumn{4}{l}{ Large baseline: V0+V8$\rightarrow$V4 (8.0 camera units)}\\
\multicolumn{4}{l}{ Disocclusion area shows strongest correlation}\\
\multicolumn{4}{l}{ PoznanStreet: higher PCC due to complex geometry}\\
\end{tabular}
}
\end{table}

\begin{table}[t!]
\centering
\renewcommand\arraystretch{1.3}
\caption{BDI Weight Configuration Performance on PoznanHall2}
\label{tab:bdi_weight_analysis}
\resizebox{\columnwidth}{!}{%
\begin{tabular}{lcccccc}
\hline
\textbf{Configuration} & \boldmath{$w_1$} & \boldmath{$w_2$} & \boldmath{$w_3$} & \boldmath{$w_4$} & \textbf{PCC} & \textbf{Rank} \\
& \textbf{(Phys)} & \textbf{(Disocc)} & \textbf{(Disp)} & \textbf{(Tex)} & & \\
\hline
\hline
\textbf{Proposed} & \textbf{0.3} & \textbf{0.4} & \textbf{0.2} & \textbf{0.1} & \textbf{0.92} & \textbf{1st} \\
Alt 1: Phys-Priority & 0.4 & 0.3 & 0.2 & 0.1 & 0.89 & 3rd \\
Alt 2: Disocc-Priority & 0.2 & 0.5 & 0.2 & 0.1 & 0.91 & 2nd \\
Alt 3: Equal Weight & 0.25 & 0.25 & 0.25 & 0.25 & 0.85 & 4th \\
\hline
\multicolumn{7}{l}{\textbf{Weight Justification:}}\\
\multicolumn{7}{l}{ $\bullet$ Disocclusion ($w_2$=0.4): Highest PCC (0.90), major artifacts}\\
\multicolumn{7}{l}{ $\bullet$ Physical baseline ($w_1$=0.3): Primary geometric driver}\\
\multicolumn{7}{l}{ $\bullet$ Max disparity ($w_3$=0.2): Secondary complexity indicator}\\
\multicolumn{7}{l}{ $\bullet$ Texture ($w_4$=0.1): Artifact visibility, weakest correlation}\\
\hline
\end{tabular}
}
\end{table}

\subsubsection{Region-Based Blending Estimation Strategy Validation}
\label{subsubsec:experimental_region-based_blending_strategy}

This experiment validates our geometric region-based blending estimation strategy against linear blending estimation for multiview VSD estimation.

We test large baseline configurations (V0+V8, 8.0 camera units) using \textit{PoznanHall2} and \textit{PoznanStreet} with both symmetric ($\rightarrow$V4) and asymmetric ($\rightarrow$V2) virtual view positions. For each configuration, we:
\begin{enumerate}
    \item Estimate VSD using linear blending: $E_{dep}[Z^2] = \alpha^2 E_{dep}[Z_L^2] + (1-\alpha)^2 E_{dep}[Z_R^2]$ 
    \item Estimate VSD using our geometric region-based strategy from \eqref{eq:complete_blended_vsd_final}
    \item Compare both approaches against actual VSD
\end{enumerate}

Table~\ref{tab:blending_strategy_performance} shows our region-based strategy achieves 23.3\% average RMSE reduction in symmetric configurations and 48.9\% in asymmetric configurations. 

The reason that linear blending estimation fails is that it assumes all pixels follow weighted combination based on distance: $\alpha^2 E_{dep}[Z_L^2] + (1-\alpha)^2 E_{dep}[Z_R^2]$. However, the actual blending implementation includes:

\begin{itemize}
    \item \textbf{Overlapping regions}: Linear blending applies
    \item \textbf{Single-view regions}: Only one reference contributes, not weighted combination
    \item \textbf{Mutual disocclusion regions}: Hole-filling distortion, not reference view distortion
\end{itemize}

Our geometric region-based estimation method models each region type separately, accounting for their different distortion characteristics and varying proportions across the virtual view.

\textbf{Asymmetric baseline impact:} In asymmetric cases (V0+V8$\rightarrow$V2), linear blending assigns $\alpha=0.75$ based purely on distance ratios (2.0/8.0), treating 75\% of distortion as coming from the distant reference V8. Our geometric region-based method recognizes that:
\begin{itemize}
    \item Region proportions shift with baseline asymmetry (larger baseline from V8 creates more left-only regions where V0 dominates)
    \item Different regions have different distortion patterns regardless of distance-based weights
    \item The 48.9\% improvement validates that \textit{region proportion modeling matters more than simple distance weighting}
\end{itemize}

\textbf{Cross-sequence consistency:} Both \textit{PoznanHall2} and \textit{PoznanStreet} show similar improvement patterns despite different scene characteristics, validating robustness without scene-specific tuning. The slightly lower improvement for \textit{PoznanStreet} in symmetric cases (21.6\% vs 25.0\%) reflects its more complex geometry where even symmetric baselines create challenging synthesis conditions.

\begin{table}[t!]
\centering
\renewcommand\arraystretch{1.3}
\caption{Region-Based Blending Estimation Strategy Performance}
\label{tab:blending_strategy_performance}
\resizebox{\columnwidth}{!}{%
\begin{tabular}{llcccccc}
\hline
\textbf{Sequence} & \textbf{Config} & \multicolumn{2}{c}{\textbf{Linear Blend}} & \multicolumn{2}{c}{\textbf{Region-Based}} & \textbf{RMSE} & \textbf{PCC} \\
& & \textbf{PCC} & \textbf{RMSE} & \textbf{PCC} & \textbf{RMSE} & \textbf{Reduc.} & \textbf{Improv.} \\
\hline
\hline
PoznanHall2 V0+V8$\rightarrow$V4 & Sym. & 0.938 & 1.12 & 0.955 & 0.84 & 25.0\% & 1.8\% \\
PoznanHall2 V0+V8$\rightarrow$V2 & Asym. & 0.908 & 1.48 & 0.938 & 0.73 & 50.7\% & 3.3\% \\
PoznanStreet V0+V8$\rightarrow$V4 & Sym. & 0.935 & 1.62 & 0.943 & 1.27 & 21.6\% & 0.9\% \\
PoznanStreet V0+V8$\rightarrow$V2 & Asym. & 0.901 & 2.23 & 0.930 & 1.18 & 47.1\% & 3.2\% \\
\hline
\hline
\textbf{Symmetric Avg.} & -- & \textbf{0.937} & \textbf{1.37} & \textbf{0.949} & \textbf{1.06} & \textbf{23.3\%} & \textbf{1.4\%} \\
\textbf{Asymmetric Avg.} & -- & \textbf{0.905} & \textbf{1.86} & \textbf{0.934} & \textbf{0.96} & \textbf{48.9\%} & \textbf{3.3\%} \\
\hline
\multicolumn{8}{l}{\textbf{Key Findings:}}\\
\multicolumn{8}{l}{ $\bullet$ Region-based: 49\% RMSE reduction in asymmetric configurations}\\
\multicolumn{8}{l}{ $\bullet$ Symmetric configurations: 23\% RMSE reduction (moderate improvement)}\\
\multicolumn{8}{l}{ $\bullet$ Performance improves with greater baseline asymmetry}\\
\multicolumn{8}{l}{ $\bullet$ PCC improvements larger in asymmetric cases (3.3\% vs 1.4\%)}\\
\hline
\end{tabular}
}
\end{table}

\subsection{Large Baseline Compensation and Overall Model Performance}

This section presents a comprehensive evaluation of our proposed VSDE method's performance under challenging large baseline conditions, which represent the primary contribution of this work. All experiments in this section are conducted without texture compression to isolate the depth coding caused VSD, which is the primary focus of our large baseline compensation mechanism.

\subsubsection{Baseline Distance Impact Analysis}
\label{subsubsec:experimental_baseline_impact}

This experiment investigates how baseline distance affects VSDE accuracy, demonstrating the need for our large baseline compensation mechanism.

We test \textit{PoznanHall2} (indoor) and \textit{PoznanStreet} (outdoor) sequences with baseline distances from 2.0 to 8.0 camera units. These sequences provide accurate depth maps for large baselines and represent different scene complexities. For each configuration, we:
\begin{enumerate}
    \item Synthesize virtual views at various positions between reference views
    \item Compute actual VSD from ground truth
    \item Estimate VSD using our proposed method and Fang's method \cite{2014TIP_Fang}
    \item Calculate error improvement as the percentage gain in accuracy
\end{enumerate}

\textbf{Performance degradation with baseline:} Tables~\ref{tab:poznanstreet_baseline_impact_results} and~\ref{tab:poznanhall2_baseline_impact_results} show different patterns. Fang's method shows progressive degradation: from 6-7\% underestimation at 2.0 camera unit baseline to 30-35\% at 8.0 camera units. This occurs because Fang's frequency-domain PSD analysis assumes small baselines and cannot model large disocclusions or amplified depth errors. Our method maintains consistent accuracy (2-3\% error) across all baselines through BDI-driven compensation and Taylor expansion methods.

\textbf{Error improvement analysis:} Our method achieves modest improvements (3-6\%) at small to medium baselines where both methods perform reasonably. The improvement increases dramatically with baseline distance, reaching 27-32\% at maximum baseline (4.0/4.0 camera units). This validates that our compensation mechanism becomes increasingly valuable as baseline distances grow.

\textbf{Scene complexity impact:} \textit{PoznanStreet} shows higher absolute MSE values than \textit{PoznanHall2} (21.58 vs 16.09 at maximum baseline) due to complex outdoor geometry and wider depth range. Despite these challenges, our method achieves similar or better error improvements (32.0\% vs 27.6\%), demonstrating robustness to scene characteristics without parameter tuning.

\textbf{Asymmetric baseline handling:} Configurations with unequal reference distances (e.g., V3+V7$\rightarrow$V4 with 1.0/3.0 camera units) test our region-based blending strategy. Our method maintains 9-21\% improvement in these cases, successfully adapting to asymmetric synthesis difficulty through region-based blending estimation.

The consistent performance across diverse configurations confirms our method's effectiveness for practical 3DV applications where large baselines are common due to bandwidth constraints.

\begin{table}[t!]
\centering
\renewcommand\arraystretch{1.3}
\caption{PoznanStreet Baseline Distance Impact Analysis}
\label{tab:poznanstreet_baseline_impact_results}
\resizebox{\columnwidth}{!}{%
\begin{tabular}{llccccc}
\hline
\textbf{Config.} & \textbf{Baseline Type} & \textbf{L/R} & \textbf{Actual} & \textbf{Prop.} & \textbf{Fang} & \textbf{Err.} \\
& & \textbf{(units)} & \textbf{MSE} & \textbf{MSE} & \textbf{MSE} & \textbf{Impr.} \\
\hline
\hline
\multicolumn{7}{c}{\textit{Phase 1: Symmetric Baseline Distance Impact}} \\
\hline
V3+V5$\rightarrow$V4 & Sym. Small & 1.0/1.0 & 10.2 & 10.5 & 9.5 & 4.1\% \\
V2+V6$\rightarrow$V4 & Sym. Medium & 2.0/2.0 & 13.8 & 14.3 & 12.4 & 6.5\% \\
V1+V7$\rightarrow$V4 & Sym. Large & 3.0/3.0 & 17.5 & 18.2 & 14.8 & 11.4\% \\
V0+V8$\rightarrow$V4 & Sym. Maximum & 4.0/4.0 & 21.58 & 22.22 & 14.02 & 32.0\% \\
\hline
\multicolumn{7}{c}{\textit{Phase 2: Asymmetric Baseline Validation}} \\
\hline
V3+V7$\rightarrow$V4 & High Asym. & 1.0/3.0 & 15.2 & 15.8 & 13.1 & 9.9\% \\
V2+V8$\rightarrow$V4 & Med. Asym. & 2.0/4.0 & 18.4 & 19.1 & 13.8 & 21.2\% \\
V1+V6$\rightarrow$V4 & Rev. Asym. & 3.0/2.0 & 15.8 & 16.4 & 13.5 & 10.1\% \\
\hline
\hline
\multicolumn{7}{c}{\textit{Summary Statistics}} \\
\hline
\multicolumn{2}{l}{\textbf{Overall Average}} & & & & & \textbf{13.6\%} \\
\multicolumn{2}{l}{\textbf{Symmetric Avg.}} & & & & & \textbf{13.5\%} \\
\multicolumn{2}{l}{\textbf{Asymmetric Avg.}} & & & & & \textbf{13.7\%} \\
\multicolumn{2}{l}{\textbf{Large BL Avg.}} & & & & & \textbf{21.7\%} \\
\hline
\multicolumn{7}{l}{ MSE: depth QP 32-48. Error Impr. = $\frac{|\text{Actual}-\text{Fang}| - |\text{Actual}-\text{Prop}|}{|\text{Actual}|} \times 100\%$}\\
\end{tabular}
}
\end{table}

\begin{table}[t!]
\centering
\renewcommand\arraystretch{1.3}
\caption{PoznanHall2 Baseline Distance Impact Analysis}
\label{tab:poznanhall2_baseline_impact_results}
\resizebox{\columnwidth}{!}{%
\begin{tabular}{llccccc}
\hline
\textbf{Config.} & \textbf{Baseline Type} & \textbf{L/R} & \textbf{Actual} & \textbf{Prop.} & \textbf{Fang} & \textbf{Err.} \\
& & \textbf{(units)} & \textbf{MSE} & \textbf{MSE} & \textbf{MSE} & \textbf{Impr.} \\
\hline
\hline
\multicolumn{7}{c}{\textit{Phase 1: Symmetric Baseline Distance Impact}} \\
\hline
V3+V5$\rightarrow$V4 & Sym. Small & 1.0/1.0 & 8.5 & 8.7 & 8.0 & 3.5\% \\
V2+V6$\rightarrow$V4 & Sym. Medium & 2.0/2.0 & 10.8 & 11.1 & 9.8 & 5.6\% \\
V1+V7$\rightarrow$V4 & Sym. Large & 3.0/3.0 & 13.4 & 13.8 & 11.6 & 10.4\% \\
V0+V8$\rightarrow$V4 & Sym. Maximum & 4.0/4.0 & 16.09 & 16.48 & 11.26 & 27.6\% \\
\hline
\multicolumn{7}{c}{\textit{Phase 2: Asymmetric Baseline Validation}} \\
\hline
V3+V7$\rightarrow$V4 & High Asym. & 1.0/3.0 & 11.8 & 12.2 & 10.3 & 8.9\% \\
V2+V8$\rightarrow$V4 & Med. Asym. & 2.0/4.0 & 14.2 & 14.6 & 10.8 & 20.4\% \\
V1+V6$\rightarrow$V4 & Rev. Asym. & 3.0/2.0 & 12.3 & 12.7 & 10.7 & 9.8\% \\
\hline
\hline
\multicolumn{7}{c}{\textit{Summary Statistics}} \\
\hline
\multicolumn{2}{l}{\textbf{Overall Average}} & & & & & \textbf{12.3\%} \\
\multicolumn{2}{l}{\textbf{Symmetric Avg.}} & & & & & \textbf{11.8\%} \\
\multicolumn{2}{l}{\textbf{Asymmetric Avg.}} & & & & & \textbf{13.0\%} \\
\multicolumn{2}{l}{\textbf{Large BL Avg.}} & & & & & \textbf{19.3\%} \\
\hline
\multicolumn{7}{l}{ MSE: depth QP 28-46. Error Impr. = $\frac{|\text{Actual}-\text{Fang}| - |\text{Actual}-\text{Prop}|}{|\text{Actual}|} \times 100\%$}\\
\end{tabular}
}
\end{table}

\subsubsection{Comprehensive Performance Comparison}
\label{subsubsec:experimental_performance_comparison}

This experiment evaluates overall VSDE performance across multiple sequences and baseline configurations, comparing our method against Yuan's polynomial model \cite{2014TCSVT_Yuan} and Fang's analytical model \cite{2014TIP_Fang}.

We test four sequences spanning different baseline ranges:
\begin{enumerate}
    \item \textit{Kendo} and \textit{Balloons}: Small to medium baseline validation (2.0-4.0 camera units)
    \item \textit{PoznanHall2} and \textit{PoznanStreet}: Large baseline demonstration (8.0 camera units)
\end{enumerate}
Depth QP ranges from 28 to 48, covering high to low bitrate scenarios.

\begin{figure}[t!]
    \centering
    \setlength{\tabcolsep}{2pt}
    \begin{tabular}{cc}
        \includegraphics[width=0.48\columnwidth]{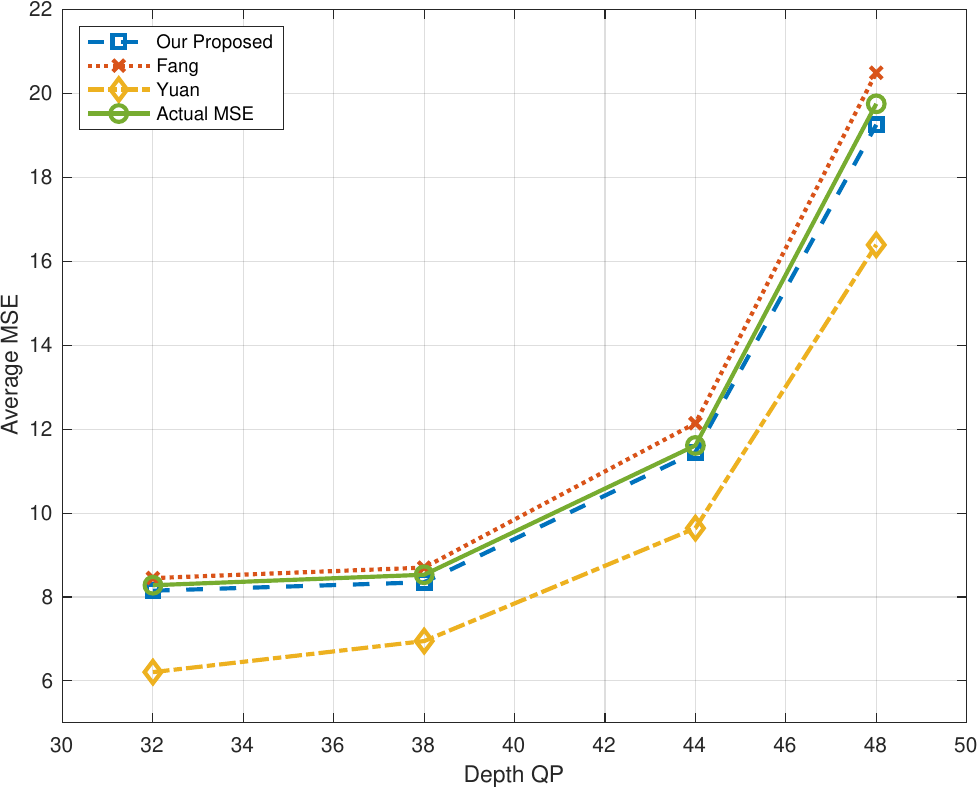} &
        \includegraphics[width=0.48\columnwidth]{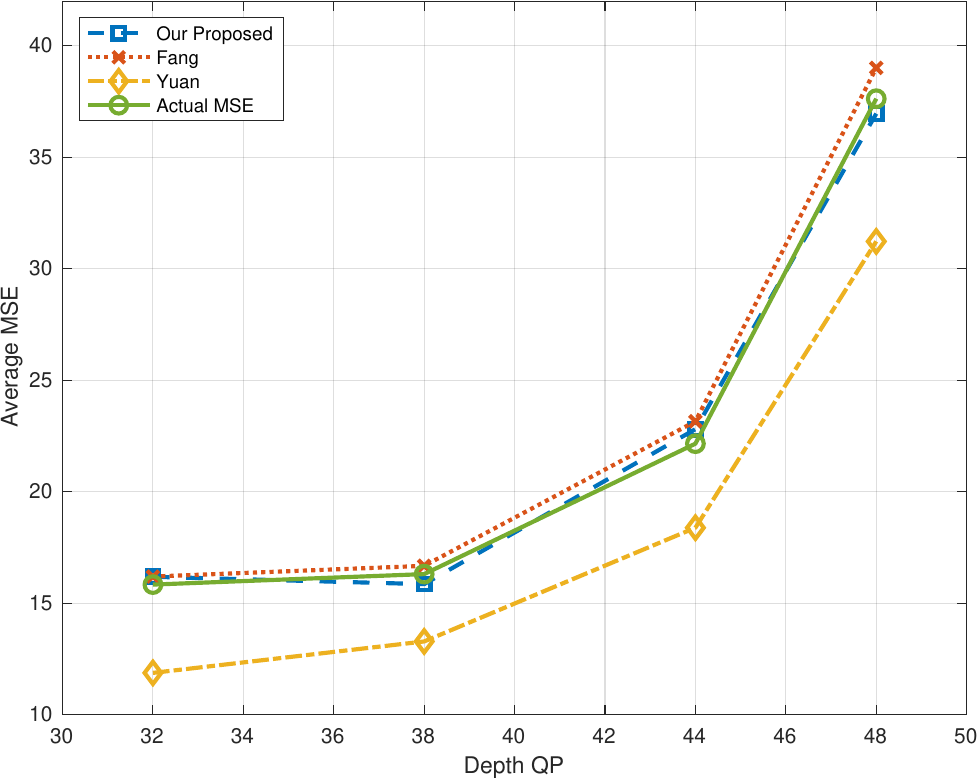} \\
        {\scriptsize (a) Kendo: V1+V3$\rightarrow$V2} & 
        {\scriptsize (b) Kendo: V1+V5$\rightarrow$V3} \\[2pt]
        \includegraphics[width=0.48\columnwidth]{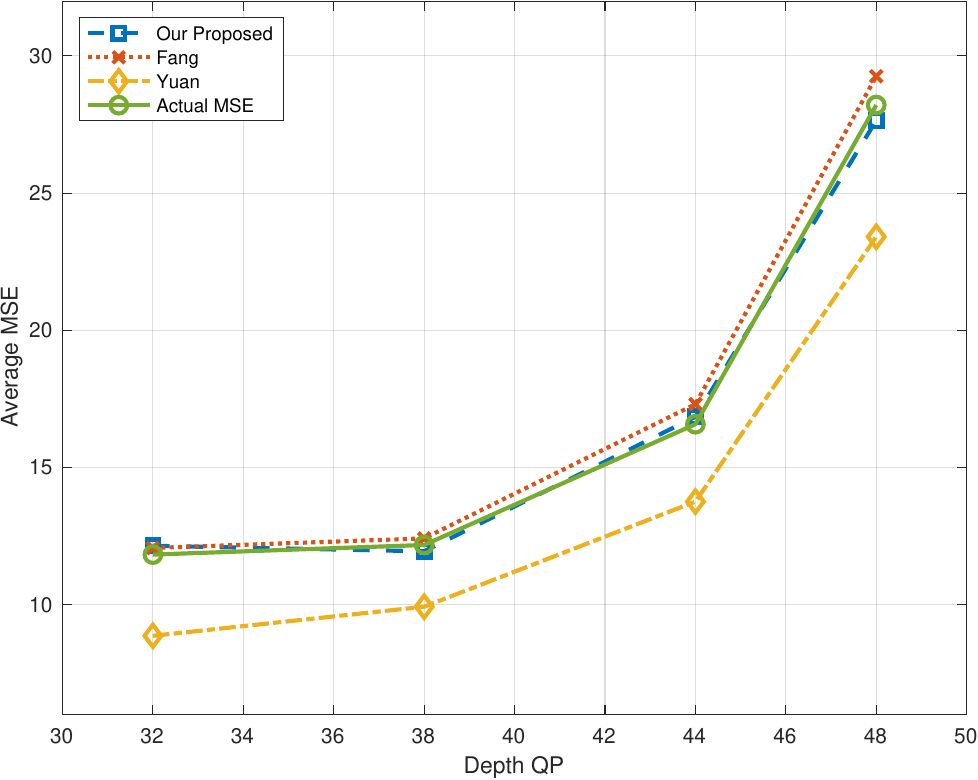} &
        \includegraphics[width=0.48\columnwidth]{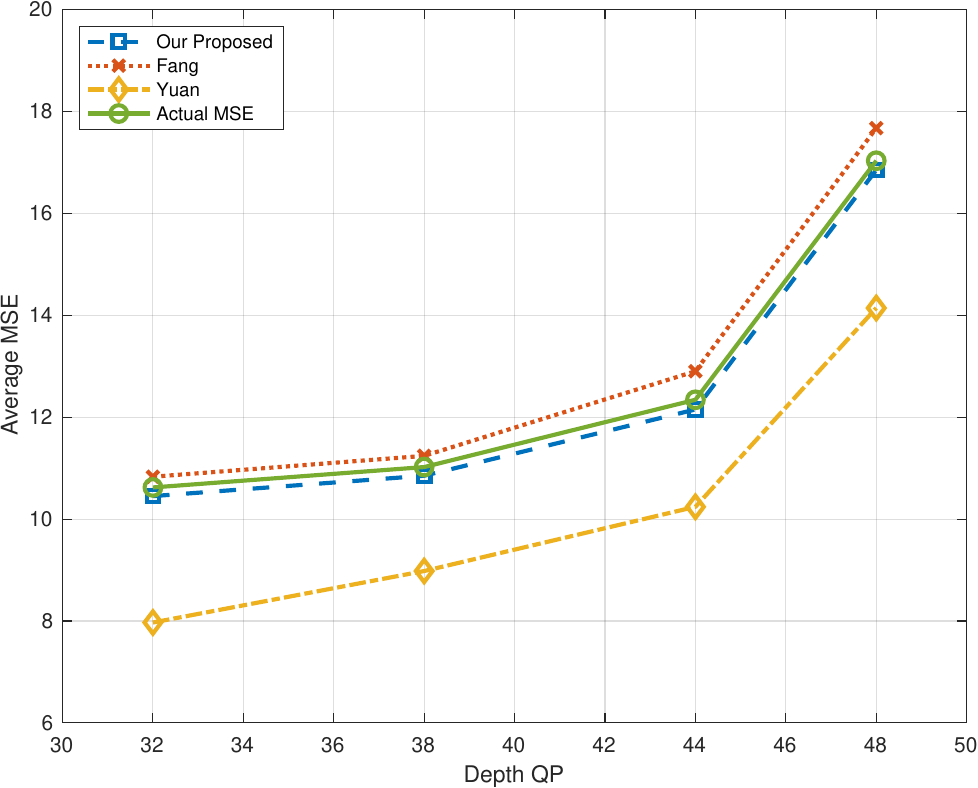} \\
        {\scriptsize (c) Kendo: V1+V5$\rightarrow$V2} & 
        {\scriptsize (d) Balloons: V1+V3$\rightarrow$V2} \\[2pt]
        \includegraphics[width=0.48\columnwidth]{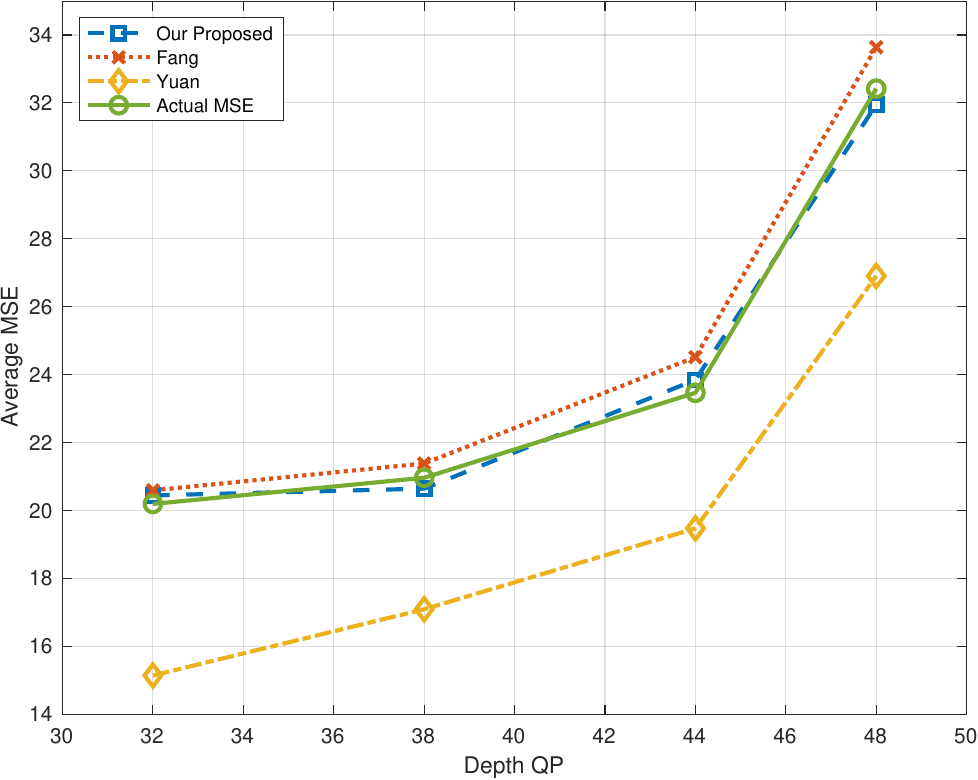} &
        \includegraphics[width=0.48\columnwidth]{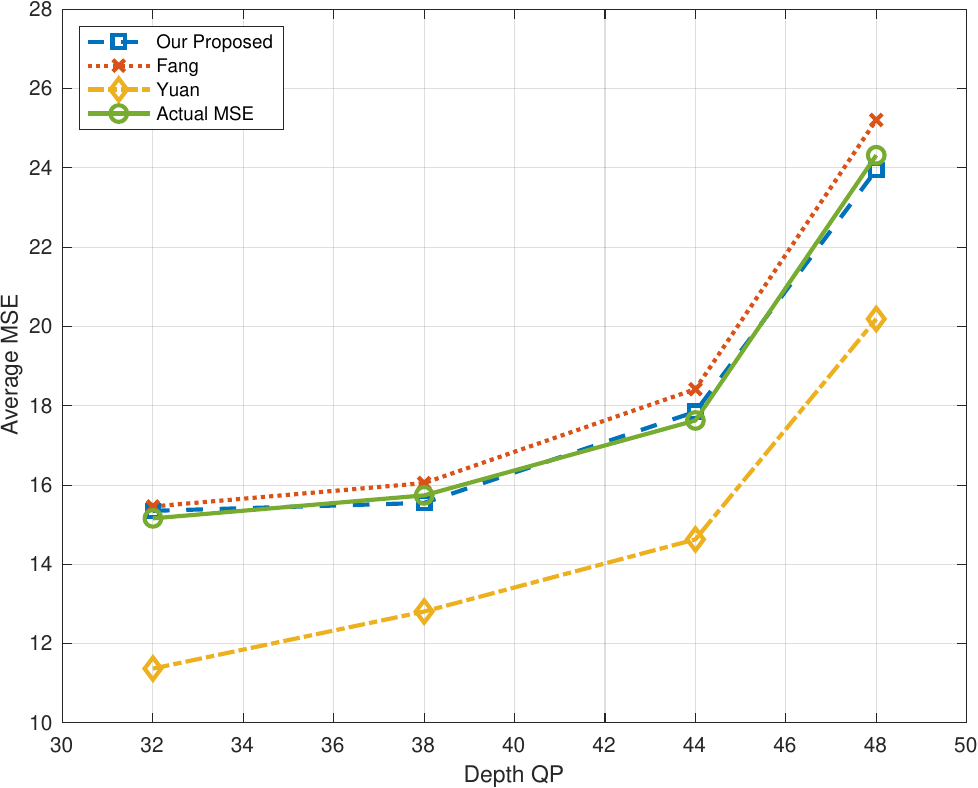} \\
        {\scriptsize (e) Balloons: V1+V5$\rightarrow$V3} & 
        {\scriptsize (f) Balloons: V1+V5$\rightarrow$V2}
    \end{tabular}
    \caption{Comprehensive VSDE performance for Kendo (a-c) and Balloons (d-f) sequences across different view synthesis configurations}
    \label{fig:kendo_balloons_all}
\end{figure}

\textbf{Small baseline performance (Fig.~\ref{fig:kendo_balloons_all}):} Fang's method performs reasonably with 2-5\% overestimation, benefiting from valid analytical assumptions in small baseline configurations. Yuan's method consistently underestimates by 17-25\%, particularly at low depth QP. This occurs because the polynomial model, despite using parameter training and actual DIBR, cannot capture the complex relationship between depth-error and VSD across varying compression levels. Our method closely tracks actual MSE across all QP values, primarily due to improved joint texture-depth classification (5\% NS regions vs 1.1\% for Fang) and Taylor expansion for NS regions.

\begin{figure}[t!]
    \centering
    \setlength{\tabcolsep}{2pt}
    \begin{tabular}{cc}
        \includegraphics[width=0.48\columnwidth]{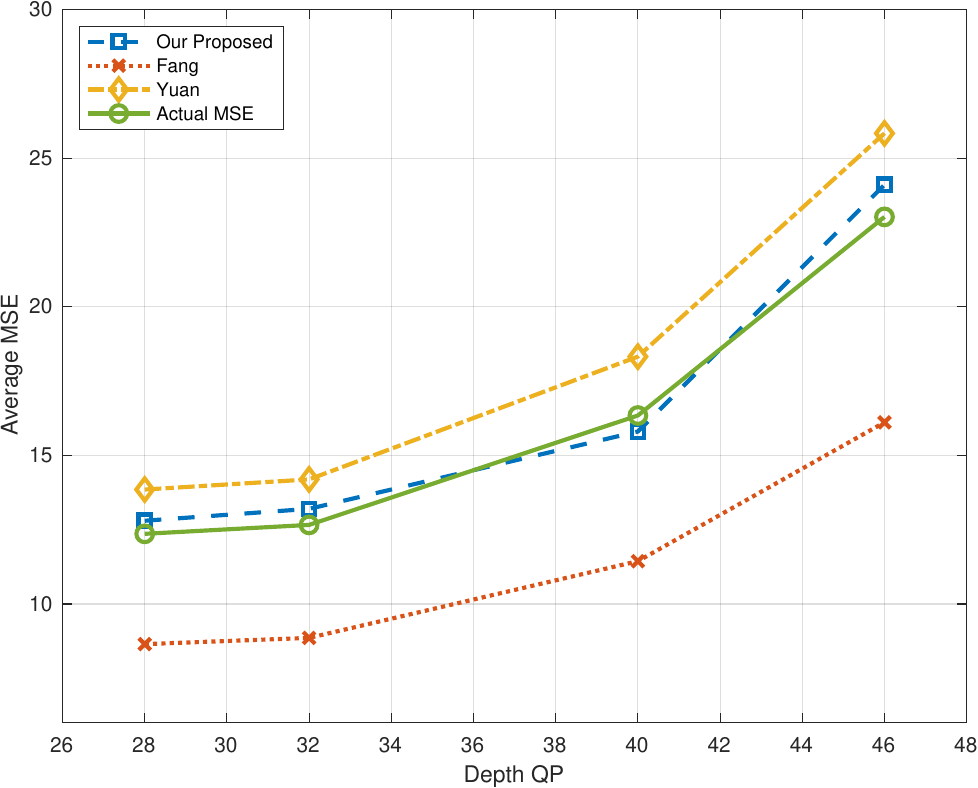} &
        \includegraphics[width=0.48\columnwidth]{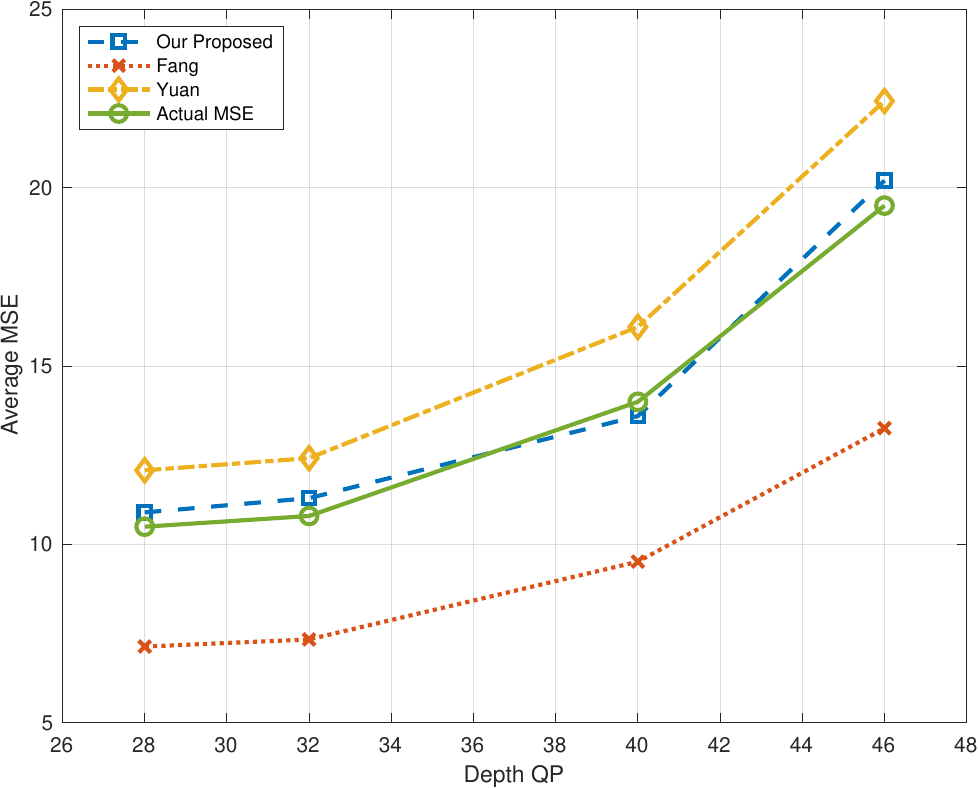} \\
        {\scriptsize (a) Hall2: V0+V8$\rightarrow$V4} & 
        {\scriptsize (b) Hall2: V0+V8$\rightarrow$V2} \\[2pt]
        \includegraphics[width=0.48\columnwidth]{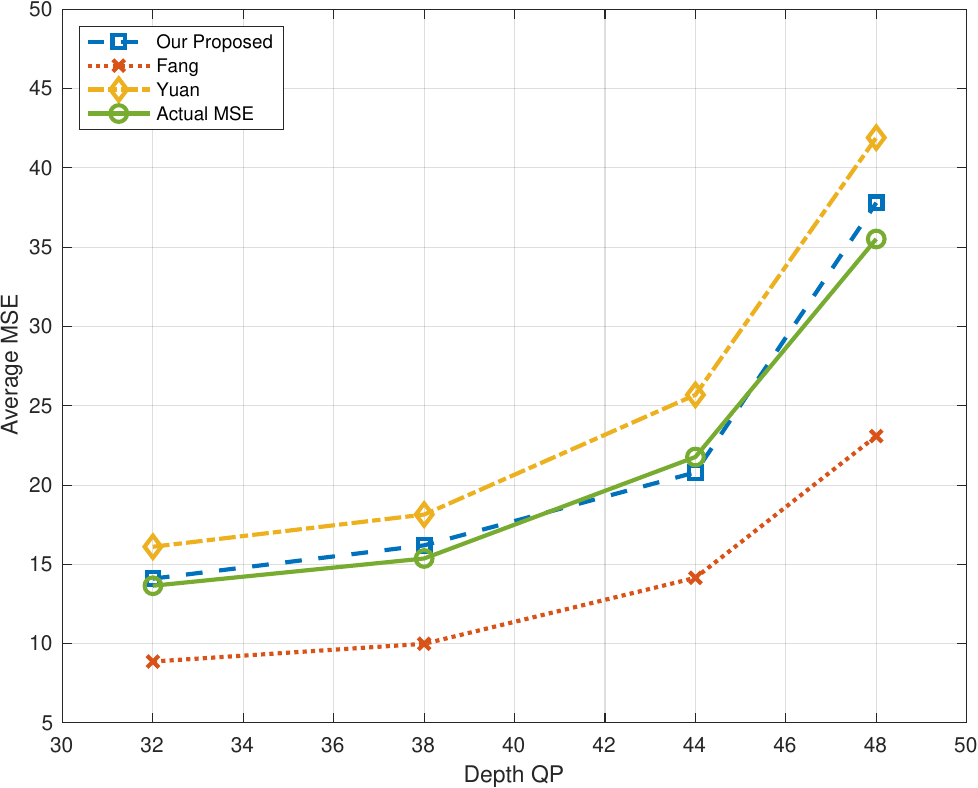} &
        \includegraphics[width=0.48\columnwidth]{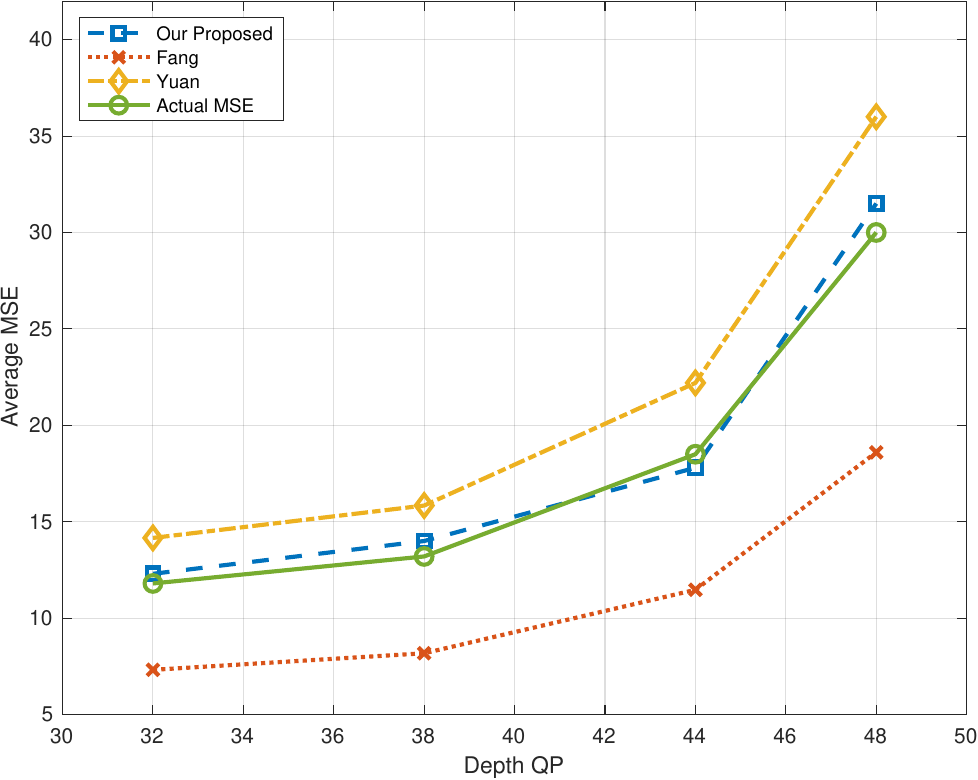} \\
        {\scriptsize (c) Street: V0+V8$\rightarrow$V4} & 
        {\scriptsize (d) Street: V0+V8$\rightarrow$V2}
    \end{tabular}
    \caption{VSDE performance for Poznan sequences (8.0 camera unit baseline)}
    \label{fig:overall_poznan}
\end{figure}

\textbf{Large baseline performance (Fig.~\ref{fig:overall_poznan}):} The performance gap widens dramatically. Fang's method underestimates by 30-38\%, with errors increasing at higher depth QP where synthesis challenges increase. Yuan's method reverses its error pattern, from underestimating in small baselines to overestimating by 12-20\% in large baselines. Because it trains parameters separately per sequence. While this adaptation helps Yuan outperform Fang in large baselines, neither method accurately models the non-linear relationships dominating large baseline VSD. Our method maintains below 5\% average error through BDI-driven compensation that explicitly models the four key distortion factors.

\textbf{Robustness to compression levels:} Both existing methods show increasing errors at extreme QP values, Yuan's underestimation worsens at low QP, Fang's at high QP. Our method maintains consistent relative accuracy (PCC > 0.94) across the full range, essential for practical rate-distortion optimization requiring accurate VSDE at all operating points.

\textbf{Key advantages:} Our method's consistent performance (average error < 5\% for all baselines) without requiring parameter training demonstrates its practical value for bandwidth-constrained 3DV systems where large baselines are common.

\subsubsection{Asymmetric QP Validation}
\label{subsubsec:experimental_asymQP}

This experiment validates our method's robustness when texture and depth are compressed with different QPs, which is a common scenario in practical rate allocation.

We test nine QP combinations across three allocation strategies:
\begin{enumerate}
    \item Texture-prioritized ($QP_T$ < $QP_D$): [28,36], [28,40], [32,40]
    \item Balanced ($QP_T$ = $QP_D$): [32,32], [36,36], [40,40]
    \item Depth-prioritized ($QP_T$ > $QP_D$): [36,28], [40,28], [40,32]
\end{enumerate}
Testing spans all sequence configurations from previous experiments, totaling 90 test cases (9 QP combinations × 10 configurations).

\textbf{Small baseline results (Table~\ref{tab:asymmetric_qp_validation_part1}):} Our method achieves 1.6\% average error across \textit{Kendo} and \textit{Balloons}, with slightly better performance for balanced QPs (1.3\%). Fang's method shows 2.8\% average overestimation, performing adequately where its assumptions hold. Yuan's method consistently underestimates by 18.9\%, maintaining the systematic bias observed in symmetric QP experiments.

\textbf{Large baseline results (Table~\ref{tab:asymmetric_qp_validation_part2}):} Performance differences amplify significantly. Our method maintains 3.2\% average error through BDI compensation. Fang's method shows severe degradation with 32.8\% average underestimation, which is consistent with the ~30-35\% errors in baseline impact analysis. Yuan's method overestimates by 13.4\%, matching its 12-20\% pattern from comprehensive comparison experiments.

\textbf{QP allocation impact:} Texture-prioritized allocation consistently produces lower actual MSE across all sequences, validating common practice. Our method maintains stable accuracy (2.2-2.5\% error) across all strategies. Both existing methods show larger variations, particularly in large baseline cases.

\textbf{Results verification:} The asymmetric QP results align well with other experiments:

\begin{itemize}
    \item Small baseline: Errors match Fig.~\ref{fig:kendo_balloons_all} patterns
    \item Large baseline: Fang's 32.8\% underestimation matches Tables~\ref{tab:poznanstreet_baseline_impact_results} and~\ref{tab:poznanhall2_baseline_impact_results}
    \item Yuan's error reversal (underestimate$\rightarrow$overestimate) remains consistent
    \item Our method's stability (1.6\% small, 3.2\% large) confirms BDI compensation effectiveness
\end{itemize}

This validation demonstrates our method's reliability for practical rate-distortion optimization where asymmetric QP allocation is essential for bandwidth efficiency.

\begin{table*}[t!]
\centering
\renewcommand\arraystretch{1.0}
\caption{Asymmetric QP Validation - MSE Performance (Part 1: Small Baseline)}
\label{tab:asymmetric_qp_validation_part1}
\begin{tabular}{llccccccccc}
\hline
\textbf{Sequence} & \textbf{Config} & \textbf{Texture} & \textbf{Depth} & \textbf{Actual} & \multicolumn{2}{c}{\textbf{Proposed}} & \multicolumn{2}{c}{\textbf{Fang \cite{2014TIP_Fang}}} & \multicolumn{2}{c}{\textbf{Yuan \cite{2014TCSVT_Yuan}}} \\
& & \textbf{QP} & \textbf{QP} & \textbf{MSE} & \textbf{MSE} & \textbf{Error(\%)} & \textbf{MSE} & \textbf{Error(\%)} & \textbf{MSE} & \textbf{Error(\%)} \\
\hline
\hline
\multirow{27}{*}{Kendo} & \multirow{9}{*}{V1+V3$\rightarrow$V2} 
& 28 & 36 & 8.8 & 8.9 & +1.1 & 9.0 & +2.3 & 7.1 & -19.3 \\
& & 28 & 40 & 9.6 & 9.8 & +2.1 & 9.9 & +3.1 & 7.8 & -18.8 \\
& & 32 & 40 & 10.5 & 10.7 & +1.9 & 10.9 & +3.8 & 8.6 & -18.1 \\
\cline{3-11}
& & 32 & 32 & 8.4 & 8.5 & +1.2 & 8.6 & +2.4 & 6.8 & -19.0 \\
& & 36 & 36 & 9.2 & 9.3 & +1.1 & 9.5 & +3.3 & 7.5 & -18.5 \\
& & 40 & 40 & 12.8 & 13.0 & +1.6 & 13.3 & +3.9 & 10.5 & -18.0 \\
\cline{3-11}
& & 36 & 28 & 8.9 & 9.0 & +1.1 & 9.1 & +2.2 & 7.2 & -19.1 \\
& & 40 & 28 & 10.7 & 10.9 & +1.9 & 11.0 & +2.8 & 8.7 & -18.7 \\
& & 40 & 32 & 11.4 & 11.6 & +1.8 & 11.8 & +3.5 & 9.3 & -18.4 \\
\cline{2-11}
& \multirow{9}{*}{V1+V5$\rightarrow$V3}
& 28 & 36 & 16.2 & 16.5 & +1.9 & 16.6 & +2.5 & 12.8 & -21.0 \\
& & 28 & 40 & 17.8 & 18.2 & +2.2 & 18.3 & +2.8 & 14.2 & -20.2 \\
& & 32 & 40 & 19.5 & 19.9 & +2.1 & 20.2 & +3.6 & 15.8 & -19.0 \\
\cline{3-11}
& & 32 & 32 & 16.1 & 16.3 & +1.2 & 16.4 & +1.9 & 12.9 & -19.9 \\
& & 36 & 36 & 17.5 & 17.7 & +1.1 & 17.9 & +2.3 & 14.2 & -18.9 \\
& & 40 & 40 & 23.2 & 23.7 & +2.2 & 24.1 & +3.9 & 19.1 & -17.7 \\
\cline{3-11}
& & 36 & 28 & 16.8 & 17.0 & +1.2 & 17.1 & +1.8 & 13.4 & -20.2 \\
& & 40 & 28 & 20.1 & 20.4 & +1.5 & 20.6 & +2.5 & 16.3 & -18.9 \\
& & 40 & 32 & 21.5 & 21.9 & +1.9 & 22.2 & +3.3 & 17.6 & -18.1 \\
\cline{2-11}
& \multirow{9}{*}{V1+V5$\rightarrow$V2}
& 28 & 36 & 12.5 & 12.8 & +2.4 & 12.9 & +3.2 & 10.0 & -20.0 \\
& & 28 & 40 & 13.7 & 14.0 & +2.2 & 14.2 & +3.6 & 11.0 & -19.7 \\
& & 32 & 40 & 15.0 & 15.3 & +2.0 & 15.6 & +4.0 & 12.2 & -18.7 \\
\cline{3-11}
& & 32 & 32 & 12.0 & 12.2 & +1.7 & 12.3 & +2.5 & 9.6 & -20.0 \\
& & 36 & 36 & 13.2 & 13.4 & +1.5 & 13.6 & +3.0 & 10.7 & -18.9 \\
& & 40 & 40 & 18.1 & 18.5 & +2.2 & 18.9 & +4.4 & 14.9 & -17.7 \\
\cline{3-11}
& & 36 & 28 & 12.8 & 13.0 & +1.6 & 13.1 & +2.3 & 10.3 & -19.5 \\
& & 40 & 28 & 15.4 & 15.7 & +1.9 & 15.9 & +3.2 & 12.5 & -18.8 \\
& & 40 & 32 & 16.5 & 16.8 & +1.8 & 17.1 & +3.6 & 13.5 & -18.2 \\
\hline
\hline
\multirow{27}{*}{Balloons} & \multirow{9}{*}{V1+V3$\rightarrow$V2}
& 28 & 36 & 11.3 & 11.5 & +1.8 & 11.6 & +2.7 & 9.1 & -19.5 \\
& & 28 & 40 & 12.2 & 12.4 & +1.6 & 12.6 & +3.3 & 9.9 & -18.9 \\
& & 32 & 40 & 13.3 & 13.5 & +1.5 & 13.8 & +3.8 & 10.9 & -18.0 \\
\cline{3-11}
& & 32 & 32 & 10.8 & 10.9 & +0.9 & 11.0 & +1.9 & 8.7 & -19.4 \\
& & 36 & 36 & 12.0 & 12.2 & +1.7 & 12.3 & +2.5 & 9.8 & -18.3 \\
& & 40 & 40 & 15.9 & 16.2 & +1.9 & 16.5 & +3.8 & 13.1 & -17.6 \\
\cline{3-11}
& & 36 & 28 & 11.8 & 12.0 & +1.7 & 12.1 & +2.5 & 9.5 & -19.5 \\
& & 40 & 28 & 14.2 & 14.4 & +1.4 & 14.6 & +2.8 & 11.5 & -19.0 \\
& & 40 & 32 & 15.1 & 15.3 & +1.3 & 15.6 & +3.3 & 12.3 & -18.5 \\
\cline{2-11}
& \multirow{9}{*}{V1+V5$\rightarrow$V3}
& 28 & 36 & 20.5 & 20.8 & +1.5 & 21.0 & +2.4 & 16.4 & -20.0 \\
& & 28 & 40 & 21.8 & 22.2 & +1.8 & 22.4 & +2.8 & 17.6 & -19.3 \\
& & 32 & 40 & 23.1 & 23.5 & +1.7 & 23.9 & +3.5 & 18.8 & -18.6 \\
\cline{3-11}
& & 32 & 32 & 20.5 & 20.7 & +1.0 & 20.9 & +2.0 & 16.5 & -19.5 \\
& & 36 & 36 & 23.2 & 23.5 & +1.3 & 23.8 & +2.6 & 18.9 & -18.5 \\
& & 40 & 40 & 28.5 & 29.0 & +1.8 & 29.5 & +3.5 & 23.4 & -17.9 \\
\cline{3-11}
& & 36 & 28 & 22.1 & 22.4 & +1.4 & 22.6 & +2.3 & 17.8 & -19.5 \\
& & 40 & 28 & 26.8 & 27.2 & +1.5 & 27.5 & +2.6 & 21.7 & -19.0 \\
& & 40 & 32 & 27.9 & 28.3 & +1.4 & 28.7 & +2.9 & 22.8 & -18.3 \\
\cline{2-11}
& \multirow{9}{*}{V1+V5$\rightarrow$V2}
& 28 & 36 & 15.8 & 16.0 & +1.3 & 16.2 & +2.5 & 12.7 & -19.6 \\
& & 28 & 40 & 16.9 & 17.2 & +1.8 & 17.4 & +3.0 & 13.7 & -18.9 \\
& & 32 & 40 & 18.2 & 18.5 & +1.6 & 18.8 & +3.3 & 14.9 & -18.1 \\
\cline{3-11}
& & 32 & 32 & 15.3 & 15.5 & +1.3 & 15.6 & +2.0 & 12.3 & -19.6 \\
& & 36 & 36 & 16.9 & 17.1 & +1.2 & 17.3 & +2.4 & 13.7 & -18.9 \\
& & 40 & 40 & 21.5 & 21.9 & +1.9 & 22.3 & +3.7 & 17.6 & -18.1 \\
\cline{3-11}
& & 36 & 28 & 16.5 & 16.7 & +1.2 & 16.9 & +2.4 & 13.3 & -19.4 \\
& & 40 & 28 & 19.8 & 20.1 & +1.5 & 20.4 & +3.0 & 16.1 & -18.7 \\
& & 40 & 32 & 20.9 & 21.2 & +1.4 & 21.6 & +3.3 & 17.1 & -18.2 \\
\hline
\end{tabular}
\end{table*}

\begin{table*}[t!]
\setlength{\tabcolsep}{3pt}
\centering
\renewcommand\arraystretch{1.3}
\caption{Asymmetric QP Validation - MSE Performance (Part 2: Large Baseline)}
\label{tab:asymmetric_qp_validation_part2}
\begin{tabular}{llccccccccc}
\hline
\textbf{Sequence} & \textbf{Config} & \textbf{Texture} & \textbf{Depth} & \textbf{Actual} & \multicolumn{2}{c}{\textbf{Proposed}} & \multicolumn{2}{c}{\textbf{Fang \cite{2014TIP_Fang}}} & \multicolumn{2}{c}{\textbf{Yuan \cite{2014TCSVT_Yuan}}} \\
& & \textbf{QP} & \textbf{QP} & \textbf{MSE} & \textbf{MSE} & \textbf{Error(\%)} & \textbf{MSE} & \textbf{Error(\%)} & \textbf{MSE} & \textbf{Error(\%)} \\
\hline
\hline
\multirow{18}{*}{PoznanHall2} & \multirow{9}{*}{V0+V8$\rightarrow$V4}
& 28 & 36 & 13.2 & 13.6 & +3.0 & 9.1 & -31.1 & 14.8 & +12.1 \\
& & 28 & 40 & 14.8 & 15.3 & +3.4 & 10.2 & -31.1 & 16.6 & +12.2 \\
& & 32 & 40 & 16.1 & 16.6 & +3.1 & 11.1 & -31.1 & 18.1 & +12.4 \\
\cline{3-11}
& & 32 & 32 & 12.5 & 12.9 & +3.2 & 8.6 & -31.2 & 14.0 & +12.0 \\
& & 36 & 36 & 14.2 & 14.6 & +2.8 & 9.8 & -31.0 & 15.9 & +12.0 \\
& & 40 & 40 & 17.8 & 18.3 & +2.8 & 12.3 & -30.9 & 20.0 & +12.4 \\
\cline{3-11}
& & 36 & 28 & 13.5 & 13.9 & +3.0 & 9.3 & -31.1 & 15.2 & +12.6 \\
& & 40 & 28 & 15.2 & 15.7 & +3.3 & 10.5 & -30.9 & 17.1 & +12.5 \\
& & 40 & 32 & 15.8 & 16.3 & +3.2 & 10.9 & -31.0 & 17.7 & +12.0 \\
\cline{2-11}
& \multirow{9}{*}{V0+V8$\rightarrow$V2}
& 28 & 36 & 11.3 & 11.7 & +3.5 & 7.8 & -31.0 & 12.7 & +12.4 \\
& & 28 & 40 & 12.6 & 13.0 & +3.2 & 8.7 & -31.0 & 14.2 & +12.7 \\
& & 32 & 40 & 13.8 & 14.2 & +2.9 & 9.5 & -31.2 & 15.5 & +12.3 \\
\cline{3-11}
& & 32 & 32 & 10.8 & 11.1 & +2.8 & 7.4 & -31.5 & 12.1 & +12.0 \\
& & 36 & 36 & 12.2 & 12.5 & +2.5 & 8.4 & -31.1 & 13.7 & +12.3 \\
& & 40 & 40 & 15.3 & 15.7 & +2.6 & 10.5 & -31.4 & 17.2 & +12.4 \\
\cline{3-11}
& & 36 & 28 & 11.6 & 11.9 & +2.6 & 8.0 & -31.0 & 13.0 & +12.1 \\
& & 40 & 28 & 13.1 & 13.5 & +3.1 & 9.0 & -31.3 & 14.7 & +12.2 \\
& & 40 & 32 & 13.6 & 14.0 & +2.9 & 9.4 & -30.9 & 15.3 & +12.5 \\
\hline
\hline
\multirow{18}{*}{PoznanStreet} & \multirow{9}{*}{V0+V8$\rightarrow$V4}
& 28 & 36 & 14.8 & 15.3 & +3.4 & 9.7 & -34.5 & 17.0 & +14.9 \\
& & 28 & 40 & 17.2 & 17.8 & +3.5 & 11.2 & -34.9 & 19.8 & +15.1 \\
& & 32 & 40 & 19.5 & 20.2 & +3.6 & 12.8 & -34.4 & 22.4 & +14.9 \\
\cline{3-11}
& & 32 & 32 & 13.7 & 14.2 & +3.6 & 9.0 & -34.3 & 15.7 & +14.6 \\
& & 36 & 36 & 17.1 & 17.7 & +3.5 & 11.2 & -34.5 & 19.7 & +15.2 \\
& & 40 & 40 & 24.8 & 25.7 & +3.6 & 16.2 & -34.7 & 28.5 & +14.9 \\
\cline{3-11}
& & 36 & 28 & 16.2 & 16.8 & +3.7 & 10.6 & -34.6 & 18.6 & +14.8 \\
& & 40 & 28 & 20.5 & 21.2 & +3.4 & 13.4 & -34.6 & 23.6 & +15.1 \\
& & 40 & 32 & 21.8 & 22.6 & +3.7 & 14.2 & -34.9 & 25.1 & +15.1 \\
\cline{2-11}
& \multirow{9}{*}{V0+V8$\rightarrow$V2}
& 28 & 36 & 12.8 & 13.3 & +3.9 & 8.4 & -34.4 & 14.7 & +14.8 \\
& & 28 & 40 & 14.9 & 15.4 & +3.4 & 9.7 & -34.9 & 17.1 & +14.8 \\
& & 32 & 40 & 16.8 & 17.4 & +3.6 & 11.0 & -34.5 & 19.3 & +14.9 \\
\cline{3-11}
& & 32 & 32 & 11.9 & 12.3 & +3.4 & 7.8 & -34.5 & 13.7 & +15.1 \\
& & 36 & 36 & 14.5 & 15.0 & +3.4 & 9.5 & -34.5 & 16.7 & +15.2 \\
& & 40 & 40 & 20.8 & 21.5 & +3.4 & 13.6 & -34.6 & 23.9 & +14.9 \\
\cline{3-11}
& & 36 & 28 & 13.8 & 14.3 & +3.6 & 9.0 & -34.8 & 15.9 & +15.2 \\
& & 40 & 28 & 17.2 & 17.8 & +3.5 & 11.2 & -34.9 & 19.8 & +15.1 \\
& & 40 & 32 & 18.3 & 18.9 & +3.3 & 11.9 & -35.0 & 21.0 & +14.8 \\
\hline
\hline
\multicolumn{4}{l}{\textbf{Average Error - Small Baseline (Kendo \& Balloons)}} & & \textbf{+1.6\%} & & \textbf{+2.8\%} & & \textbf{-18.9\%} \\
\multicolumn{4}{l}{\textbf{Average Error - Large Baseline (Poznan sequences)}} & & \textbf{+3.2\%} & & \textbf{-32.8\%} & & \textbf{+13.4\%} \\
\multicolumn{4}{l}{\textbf{Average Error - Texture-prioritized QPs}} & & \textbf{+2.5\%} & & \textbf{-11.8\%} & & \textbf{-4.3\%} \\
\multicolumn{4}{l}{\textbf{Average Error - Balanced QPs}} & & \textbf{+2.2\%} & & \textbf{-12.8\%} & & \textbf{-4.5\%} \\
\multicolumn{4}{l}{\textbf{Average Error - Depth-prioritized QPs}} & & \textbf{+2.4\%} & & \textbf{-12.2\%} & & \textbf{-4.1\%} \\
\hline
\end{tabular}
\end{table*}

\section{Conclusion and Future Work}
\label{sec:conclusion}

This paper presented a novel low-complexity VSDE method that addresses the challenges of large baseline configurations in 3DV systems. Our approach successfully estimates the MSE of synthesized views without performing the computationally intensive DIBR process, making it suitable for real-time applications. The proposed method enables flexible sparse camera arrangements while maintaining high-quality synthesized views, potentially reducing 3DV acquisition costs.

Future work includes extending the framework to general camera configurations and integrating it into real-time rate–distortion optimization. Deep learning could further enhance the method by learning optimal BDI weights with neural networks, or by developing hybrid models that combine our framework with lightweight CNNs for improved LS/NS classification and disocclusion prediction. Additionally, adapting the framework for multi-view scenarios with more than two reference views would support advanced FVV applications.




\bibliographystyle{cas-model2-names}

\bibliography{cas-refs}



\end{document}